\documentclass[notitlepage,amsmath,superscriptaddress,amssymb,11pt,floatfix,aps,prd,nofootinbib,preprintnumbers]{revtex4-2}

\usepackage{setspace}
\usepackage{amssymb,amsfonts,multirow}

\usepackage{orcidlink}
\usepackage{amssymb,url,bm}
\usepackage{graphicx}
\usepackage{ragged2e}
\usepackage{hyperref}
\usepackage{color}

\newcommand{\SU}{\mathrm{SU}}

\usepackage{slashed}
\usepackage[export]{adjustbox}

\newcommand{\gam}[2]{\gamma_{#1}^{(#2)}}
\long\def\del #1 \enddel { }

\usepackage{epsfig}
\usepackage{xcolor,colortbl}

\definecolor{Gray}{gray}{0.85}
\definecolor{LightGreen}{rgb}{0.88, 1, 0.88}
\definecolor{Lime}{rgb}{0,255,0}
\definecolor{LightCyan}{rgb}{0.88,1,1}
\definecolor{LightRed}{rgb}{1, 0.85, 0.85}
\definecolor{Red}{rgb}{1, 0, 0}
\definecolor{LightYellow}{rgb}{1, 1, 0.85}
\definecolor{Yellow}{rgb}{1,1,0.05}
\definecolor{LightBlue}{rgb}{0.87, 0.94, 1}
\definecolor{white}{gray}{1}
\definecolor{black}{gray}{0}
\newcolumntype{G}{>{\columncolor{LightGray}}c}

\definecolor{LightGray}{gray}{0.93}

\usepackage{epsfig,bm}

\usepackage{array,mathtools,amssymb,booktabs}
\newcolumntype{C}{>{$}c<{$}}
\AtBeginDocument{
\heavyrulewidth=.16em
\lightrulewidth=.1em
\cmidrulewidth=.03em
\belowrulesep=.4ex
\belowbottomsep=0pt
\aboverulesep=.4ex
\abovetopsep=0pt
\cmidrulesep=\doublerulesep
\cmidrulekern=.5em
\defaultaddspace=.5em
}

\usepackage{hyperref}

\renewcommand{\thesection}{{\bf \Roman{section}}}

\usepackage{hyperref}

\def\beq{\begin{equation}}
\def\eeq{\end{equation}}

\newcommand*{\bz}[1]{{BZ}$_{#1}$}
\newcommand*{\gy}[1]{{GY}$_{#1}$}
\newcommand*{\BZ}[1]{$\bm{{\rm BZ}_{#1}}$}
\newcommand*{\GY}[1]{$\bm{{\rm GY}_{#1}}$}

\def\bea{\arraycolsep .1em \begin{eqnarray}}
\def\eea{\end{eqnarray}}
\def\Tr{{\rm Tr}}

\def\eps{\epsilon}

\def\al#1{\alpha_{\rm {#1}}}
\def\eq#1{(\ref{#1})}

\def\s0#1#2{\mbox{\small{$ \frac{#1}{#2} $}}}
\def\0#1#2{\frac{#1}{#2}}

\def\al#1{\alpha_#1}
\def\grgl{\:\hbox to -0.2pt{\lower2.5pt\hbox{$\sim$}\hss}{\raise3pt\hbox{$>$}}\:}
\def\klgl{\:\hbox to -0.2pt{\lower2.5pt\hbox{$\sim$}\hss}{\raise3pt\hbox{$<$}}\:}

\newcommand{\eff}{\textrm{eff}}

\def\lsim{\mathrel{\rlap{\lower4pt\hbox{\hskip1pt$\sim$}}
 \raise1pt\hbox{$<$}}} 
\def\gsim{\mathrel{\rlap{\lower4pt\hbox{\hskip1pt$\sim$}}
 \raise1pt\hbox{$>$}}} 

\makeatletter
 \def\CT@@do@color{%
 \global\let\CT@do@color\relax
 \@tempdima\wd\z@
 \advance\@tempdima\@tempdimb
 \advance\@tempdima\@tempdimc
 \advance\@tempdimb\tabcolsep
 \advance\@tempdimc\tabcolsep
 \advance\@tempdima2\tabcolsep
 \kern-\@tempdimb
 \leaders\vrule
 \hskip\@tempdima\@plus 1fill
 \kern-\@tempdimc
 \hskip-\wd\z@ \@plus -1fill }
 \makeatother

\newcommand{\Ncal}{\mathcal{N}}
\newcommand{\und}{\quad\mathrm{and}\quad}

\newcommand{\nocontentsline}[3]{}
\let\origcontentsline\addcontentsline
\newcommand\stoptoc{\let\addcontentsline\nocontentsline}
\newcommand\resumetoc{\let\addcontentsline\origcontentsline}

\PassOptionsToPackage{unicode}{hyperref}
\PassOptionsToPackage{naturalnames}{hyperref}

\begin{document}
${}$\vskip1cm

\title{Fixed Points in \texorpdfstring{$\mathrm{SU}(N)\times \mathrm{SU}(M)$}{SU(N)xSU(M)} Supersymmetric Gauge Theories}
\title{Fixed points of semi-simple supersymmetric gauge theories}

\author{Andrew~D.~Bond\orcidlink{0000-0001-7109-9317}}
\affiliation{\mbox{Department of Physics and Astronomy, U Sussex, Brighton, BN1 9QH, U.K.}}
\author{Daniel F.~Litim\orcidlink{0000-0001-9963-5345}}
\email{d.litim@sussex.ac.uk}
\affiliation{\mbox{Department of Physics and Astronomy, U Sussex, Brighton, BN1 9QH, U.K.}}

\author{Gabriel Picanço\orcidlink{0000-0002-2698-607X}}
\email{g.picanco@sussex.ac.uk}
\affiliation{\mbox{Department of Physics and Astronomy, U Sussex, Brighton, BN1 9QH, U.K.}}
\affiliation{\mbox{Department of Physics, TU Dortmund, Otto-Hahn-Stra\ss e 4, 44227 Dortmund, Germany}}
\vskip-3cm

\begin{abstract}
We study fixed points and phase diagrams of semi-simple supersymmetric gauge theories coupled to chiral superfields and a superpotential. Particular emphasis is put on new phenomena which arise due to the semi-simple nature of gauge interactions and the constraints dictated by supersymmetry, unitarity, and the $a$-theorem. Using field multiplicities as free parameters, we find all superconformal fixed points and classify theories according to their phase diagrams. Highlights include asymptotically free theories displaying a range of interacting fixed points in the IR, asymptotically non-free theories that become asymptotically safe due to residual interactions, UV-complete theories with gauge sectors that are simultaneously UV-free and IR-free, and theories that remain interacting both in the asymptotic UV and IR. Estimates for the sizes of conformal windows are also provided, and implications for model building are discussed.
\end{abstract}

\maketitle

\begin{spacing}{.87}
\tableofcontents
\end{spacing}

\section{\bf Introduction}

Critical phenomena in quantum and statistical field theory are characterised by fixed points of the renormalisation group. Infrared (IR) critical points often relate to continuous quantum phase transitions and govern low-energy phenomena such as spontaneous symmetry breaking or the generation of mass. Ultraviolet (UV) fixed points are key for the {\it bona fide} predictivity of quantum field theory \cite{Wilson:1971bg}. The latter may be free such as in asymptotic freedom \cite{Gross:1973id, Politzer:1973fx}, or interacting, such as in asymptotically safe gauge theories with matter \cite{Bailin:1974bq,Litim:2014uca,Bond:2016dvk,Bond:2018oco}, 
fermionic theories \cite{Rosenstein:1988pt,Cresswell-Hogg:2022lgg,Cresswell-Hogg:2025wda},
or even quantum gravity \cite{Weinberg:1980gg,Reuter:2001ag,Litim:2003vp, Falls:2014tra,Fehre:2021eob,Kluth:2024lar}.

The discovery of asymptotic safety in particle physics \cite{Litim:2014uca,Bond:2016dvk,Bond:2018oco} has opened up a new door for model building  \cite{Bond:2019npq,Bond:2017lnq,Bond:2017suy,Bond:2022xvr,Hiller:2022hgt,Litim:2015iea,Bond:2017tbw,Bond:2021tgu,Litim:2023tym,Hiller:2019tvg,Hiller:2019mou,Hiller:2020fbu,Bause:2021prv,Hiller:2022rla,Bissmann:2020lge,Bause:2022jes,Kowalska:2017fzw,Abel:2017ujy,Buyukbese:2017ehm,Bond:2022xvr}, beyond the paradigms of asymptotic freedom or effective field  theory. 
Here, asymptotic safety arises purely as a quantum effect through subtle cancelations of fluctuations between the elementary gauge, fermion, and scalar fields. By now, general conditions for this to occur at weak coupling have been identified \cite{Bond:2016dvk,Bond:2018oco}, and UV-safe templates are known for unitary \cite{Litim:2014uca}, orthogonal, symplectic \cite{Bond:2019npq}, or product gauge groups \cite{Bond:2017lnq}, and in supersymmetry \cite{Bond:2017suy,Bond:2022xvr,Hiller:2022hgt}. Further results cover aspects of vacuum stability \cite{Litim:2015iea}, conformal windows up to four loops in perturbation theory \cite{Bond:2017tbw,Bond:2021tgu,Litim:2023tym}, Higgs vacuum stability and model building \cite{Hiller:2019tvg,Hiller:2019mou,Hiller:2020fbu,Bause:2021prv,Hiller:2022rla}, testable collider signatures \cite{Bissmann:2020lge}, explanations for new anomalies in charm \cite{Bause:2022jes}, the role of Abelian factors \cite{Kowalska:2017fzw}, aspects of radiative symmetry breaking \cite{Abel:2017ujy}, and higher-order extensions \cite{Buyukbese:2017ehm}.

In supersymmetry, it is well known that asymptotically free theories continue to display a plethora of IR critical points. The primary mechanism for asymptotic safety \cite{Litim:2014uca,Bond:2016dvk,Bond:2018oco}, however, is not operative \cite{Martin:2000cr,Intriligator:2015xxa}. Still, asymptotic safety can arise when quantum fluctuations turn marginally irrelevant interactions into marginally relevant ones, which necessitates semi-simple gauge groups \cite{Bond:2017suy}.
In practice, this makes UV-safe models with supersymmetry both more constrained and more predictive. Further, the availability of infinite-order gauge beta functions \cite{Novikov:1983uc}, powerful non-renormalisation theorems \cite{Grisaru:1979wc}, Seiberg duality \cite{SEIBERG1995129} and $a$-maximisation \cite{Intriligator:2003jj,Barnes:2004jj,Barnes:2005zn} are assets to help understand fixed points at strong coupling. As a result, the landscape of UV-safe models has been found to be significantly larger than the part visible in perturbation theory \cite{Bond:2022xvr}, and has triggered searches for UV-safe extensions of the minimal supersymmetric standard model (MSSM) \cite{Hiller:2022hgt}.

In this paper, we investigate fixed points and conformal windows in general semi-simple supersymmetric gauge theories with matter and a single superpotential coupling. Expanding upon \cite{Bond:2017lnq,Bond:2017suy,Bond:2022xvr}, we are particularly interested in new phenomena which arise due to the semi-simple nature of gauge interactions, the availability of UV conformal fixed points that may serve as templates for model building, and the constraints dictated by supersymmetry, unitarity, and the $a$-theorem. For a concrete class of models with $\SU(N)\times \SU(M)$ gauge groups, and using field multiplicities as free parameters, we 
determine all superconformal fixed points and classify theories according to their asymptotics. We also investigate the size of conformal windows, benchmark against exact results, and highlight new effects for model building.

The paper is structured as follows. We first discuss the range of interacting fixed points and scaling exponents in models with supersymmetry (Sec.~\ref{sec:FPgeneral}). For concrete templates with $\SU(N)\times \SU(M)$ gauge symmetry, we conduct a comprehensive fixed point analysis in a Veneziano large-$N$ limit, while keeping the gauge group dimensions and matter field multiplicities as free parameters (Sec.~\ref{sec:MM}). In combination, this leads to a complete classification of phase diagrams of theories according to their UV and IR asymptotics (Sec.~\ref{sec:PD}). In order to estimate the size of conformal windows, we extend the analysis to three-loop order, and benchmark against unitarity bounds and exact results (Sec.~\ref{sec:CW}). We conclude with a summary of findings and implications for model building (Sec.~\ref{sec:sum}). Two appendices contain technical material (App.~\ref{app:beta-functions-NNLO}) and results for Banks--Zaks conformal windows at three loops (App.~\ref{app:BZs}).

\section{\bf Fixed Points with Supersymmetry}\label{sec:FPgeneral}

In this section, we recall aspects of interacting fixed points in semi-simple supersymmetric gauge theories which are weakly coupled to matter, with or without a superpotential \cite{Bond:2016dvk,Bond:2017sem}. We also introduce some notation and conventions.

\subsection{Perturbation Theory}

We are interested in the renormalisation of general supersymmetric gauge theories coupled to chiral matter multiplets. The running of the gauge couplings $\alpha_i \propto g_i^2$ with the renormalisation group scale $\mu$ is determined by the beta functions of the theory. Expanding them perturbatively up to two loops, we have
\begin{align}\label{eq:RGE}
	\mu \partial_\mu \alpha_i \equiv \beta_i= \alpha_i^2(-B_i + C_{ij}\alpha_j - 2\,Y_{4,i}) + \mathcal{O}(\alpha^4)\,,
\end{align}
where a sum over gauge group factors $j$ is implied. The one- and two-loop gauge contributions $B_i$ and $C_{ij}$ and the two-loop Yukawa contributions $Y_{4,i}$ are known for general gauge theories, see~\cite{Machacek:1983tz,Machacek:1983fi,Machacek:1984zw,Luo:2002ti,Bond:2016dvk} for explicit expressions. While $B_i$ and $C_{ii}$ may take either sign, depending on the matter content, the Yukawa contribution $Y_{4,i}$ and the off-diagonal gauge contributions $C_{ij}$ $(i\neq j)$ are strictly positive in any quantum field theory. The effect of Yukawa couplings can incorporated by projecting the gauge beta functions \eq{eq:RGE} onto the Yukawa nullclines $(\beta_y = 0)$, leading to explicit expressions for $Y_{4,i}$ in terms of the gauge couplings $\alpha_j$. Moreover, for many theories, the Yukawa contribution along nullclines can be written as $Y_{4,i}=D_{ij}\,\alpha_j$, with $D_{ij} \ge 0$ \cite{Bond:2016dvk}. We can then go one step further and express the net effect of Yukawa couplings as a shift of the two-loop gauge contribution, $C_{ij} \rightarrow C_{ij}' = C_{ij} - 2D_{ij} \leq C_{ij}$. Notice that the shift will always be by some negative amount provided at least one of the Yukawa couplings is non-vanishing. It leads to the reduced gauge beta functions
\begin{align}\label{eq:RGE2}
	\beta_i = \alpha_i^2(-B_i + C'_{ij}\alpha_j) + \mathcal{O}(\alpha^4)\,.
\end{align}
Fixed points solutions of \eq{eq:RGE2} are either free or interacting and $\alpha^*=0$ for some or all gauge factors is always a self-consistent solution. Consequently, interacting fixed points are solutions to
\begin{align}\label{eq:matFP}
	B_i &= C_{ij}'\,\alpha_j^*\,,\quad\text{subject to}\quad\alpha_i^*> 0\,,
\end{align}
where only those rows and columns are retained where gauge couplings are interacting (see Tab.~\ref{tab:tFPdef} for our conventions).

\begin{table*}
\begin{center}
\begin{tabular}{cccc}
\toprule
\rowcolor{Yellow}
&&&\\[-4mm]
\rowcolor{Yellow}
\multicolumn{2}{c}{\bf Fixed Point} 
&${}\quad\bm\alpha_{\rm \bf Gauge}{}\quad$
&${}\quad\bm\alpha_{\rm \bf Yukawa}{}\quad$
\\
\midrule
\rowcolor{LightGray}
Gauss&G&$=0$&$=0$\\
Banks--Zaks&BZ& $\neq 0$&$=0$\\
\rowcolor{LightGray}
Gauge-Yukawa\ \ &GY& $\neq 0$ & $\neq 0$\\
\bottomrule
\end{tabular}
\caption{Conventions for the naming of fixed points in gauge theories coupled to matter.}
\label{tab:tFPdef}
\end{center}
\end{table*}

Next we discuss the role of superpotential (Yukawa) couplings. In the absence of Yukawa couplings, the two-loop coefficients remain unshifted, $C_{ij}' = C_{ij}$. An immediate consequence of this is that any interacting fixed point must necessarily be IR. The reason is as follows: for an interacting fixed point to be UV, asymptotic freedom cannot be maintained for all gauge factors, meaning that some $B_i < 0$. However, as has been established in \cite{Bond:2016dvk}, $B_i \leq 0$ necessarily entails $C_{ij} \geq 0$ in any $4d$ quantum gauge theory. If the left hand side of \eqref{eq:matFP} is negative, if only for a single row, positivity of $C_{ij}$ requires that some $\alpha_j^*$ must take negative values for a fixed point solution to arise. This, however, is unphysical \cite{Dyson:1952tj} and we are left with $B_i > 0$ for each $i$, implying that asymptotic freedom remains intact in all gauge sectors. Besides the Gaussian, the theory may have weakly interacting infrared Banks--Zaks fixed points in each gauge sector, as well as products thereof, which arise as solutions to \eq{eq:matFP} with the unshifted coefficients.

In the presence of Yukawa couplings, the coefficients $C_{ij}'$ can in general take either sign. This has far reaching implications. Firstly, the theory can additionally display gauge-Yukawa fixed points where both the gauge and the Yukawa couplings are non-zero. Most importanly, solutions to \eqref{eq:matFP} are then no longer limited to theories with asymptotic freedom. Instead, interacting fixed points can be infrared, ultraviolet, or of the crossover type. In general we may expect gauge-Yukawa fixed points for each independent Yukawa nullcline. In summary, perturbative fixed points are either $(i)$ free and given by the Gaussian, or $(ii)$ free in the Yukawa but interacting in the gauge sector (Banks--Zaks fixed points), or $(iii)$ simultaneously interacting in the gauge and the Yukawa sector (gauge-Yukawa fixed points), or $(iv)$ combinations and products of $(i)$, $(ii)$ and $(iii)$. Banks--Zaks fixed points are always IR, while the Gaussian and gauge-Yukawa fixed points can be either UV or IR. Depending on the details of the theory and its Yukawa structure, if the theory is not effective, either the Gaussian or one of the interacting gauge-Yukawa fixed points will arise as the ``ultimate'' UV fixed point of the theory and may serve to define the theory fundamentally \cite{Bond:2017sem}.

\subsection{Consequences of Supersymmetry}\label{subsec:susy}

Before we look into particular gauge groups and Yukawa structures, let us consider two important consequences of supersymmetry, namely a consistency condition for the existence of interacting UV or IR fixed points, and the uniqueness of fixed point types as dictated by the superconformal U$(1)_R$-symmetry.
Let's consider any $\mathcal{N}=1$ supersymmetric gauge theory with product gauge group 
\begin{equation}
 	{G} = \bigotimes_a {G}_a\,,
\end{equation}
where ${G}_a$ are simple factors with dimension $d(G_a)$, quadratic Casimir $C_2^a$, gauge couplings $\alpha_a \equiv (g_a/4\pi)^2$, and one-loop coefficients $B_a$, as in \eqref{eq:RGE2}. The theory is further coupled to chiral superfield including a superpotential. Then, for the theory to display an interacting fixed point, the presence of superpotential couplings implies a consistency condition \cite{Martin:2000cr,Hiller:2022hgt}, namely
\begin{equation}
	\sum_a B_a\,d(G_a)\,\alpha_a^* \geq 0\,,
\end{equation}
where $\alpha_a^*$ are the gauge couplings at the fixed point. Since $\alpha_a^*\geq0$ for any physical fixed points, the positivity of the sum requires at least one of the universal loop factors $B_a$ to be positive, implying that such a gauge sector would be free in the UV. This has two immediate consequences: firstly, supersymmetric theories with a single gauge sector must be asymptotically free in order to display an isolated interacting fixed point. Secondly, for a non-asymptotically free supersymmetric gauge theory to become asymptotically safe requires at least two gauge sectors, at least one of which has to remain asymptotically free. For this reason, as we are interested in the possible existence of interacting UV fixed points, throughout this paper we will work with gauge groups of the form ${\cal G}_1\otimes {\cal G}_2$, which are the simplest gauge groups compatible with asymptotic safety.

\begin{figure}[t]
\includegraphics[width=0.3\linewidth]{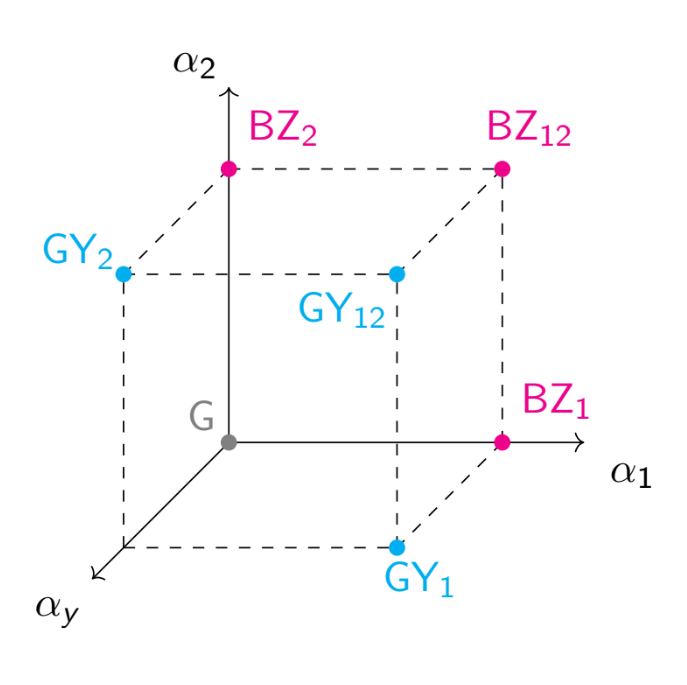}
\caption{
Schematic plot illustrating the maximal set of isolated fixed points of a supersymmetric gauge theory with two gauge and a single Yukawa coupling, showing the Gaussian fixed point (gray) and interacting fixed points of the Banks--Zaks (magenta) and gauge-Yukawa type (cyan), see Tabs.~\ref{tab:tFPdef} and~\ref{tab:tFPs}.
}
\label{fig:FP-structure}
\end{figure}

The second general result that distinguishes supersymmetric gauge theories from non-super\,sym\-metric ones relates to the $a$-theorem \cite{Zamolodchikov:1986gt,Osborn:1989td,Jack:1990eb,Cardy:1988cwa,Komargodski:2011vj,Komargodski:2011xv} and the superconformal and anomaly-free U$(1)_R$-symmetry. The latter dictates unique $R$-charges for all chiral superfields at any interacting fixed point of the theory, which can be determined using the technique of $a$-maximisation \cite{Intriligator:2003jj}. This also entails a value for the conformal anomaly $a$, which can be expressed uniquely in terms of the $R$-charges \cite{Anselmi:1997am,Anselmi:1997ys}. It follows that $R$-charges and $a$-function agree for any ``type'' of fixed point where the same set of couplings are non-zero. The $a$-theorem states that the value of the $a$-function must decrease along the RG flow from one fixed point to another. It follows that fixed points where the same set of couplings are non-zero cannot be connected by an RG flow. Hence, either there exists, at best, a single isolated fixed point of any type, or fixed points degenerate into a line of fixed points. For a theory with two gauge sectors and one Yukawa superpotential coupling, the maximally achievable set of isolated fixed points is illustrated in Fig.~\ref{fig:FP-structure}.

\subsection{Gauge Couplings}

Let us now consider a semi-simple gauge-Yukawa theory with non-Abelian gauge fields under the semi-simple gauge group ${\cal G}_1\otimes {\cal G}_2$ coupled to superfields. We have two non-Abelian gauge couplings $\al 1$ and $\al 2$, which are related to the fundamental gauge couplings via $\alpha_i=g_i^2/(4\pi)^2$. 
The running of gauge couplings within perturbation theory is given by
\beq\label{beta12}
\begin{array}{rcl}
\beta_1&=&
\displaystyle
-B_1\, \alpha_1^2+C_1\,\alpha_1^3 +G_1\,\alpha_1^2\,\alpha_2\,,\\[1ex]
\beta_2&=&
\displaystyle
 -B_2\, \alpha_2^2+C_2\,\alpha_2^3 +G_2\,\alpha_2^2\,\alpha_1\,.
\end{array}
\eeq
Here, $B_i$ are the well-known one-loop coefficients. In theories without superpotential, the numbers $C_i$ and $G_i$ are the two-loop coefficients which arise owing to the gauge loops and to the mixing between gauge groups, meaning $C_i\equiv C_{ii}$ (no sum), and $G_{1}\equiv C_{12}$, $G_{2}\equiv C_{12}$, see \eq{eq:RGE}. In this case, we also have that $C_i, G_i\ge 0$ as soon as $B_i<0$.\footnote{General formal expressions of loop coefficients in the conventions used here are given in \cite{Bond:2016dvk}.} For theories where superpotential couplings take interacting fixed points, the numbers $C_i$ and $G_i$ receive corrections, as $C_i\equiv C'_{ii}$ (no sum) and $G_{1}\equiv C'_{12}$, $G_{2}\equiv C'_{12}$, see \eq{eq:RGE2}, and strict positivity of $C_i$ and $G_i$ is not guaranteed \cite{Bond:2016dvk}.

Fixed points of the combined system are determined by the vanishing of \eq{beta12}. The Gaussian fixed point
\beq
\label{1}
(\al1^*,\al2^*)=(0,0)
\eeq
always exists (see Tab.~\ref{tab:tFPdef} for our conventions). It is the UV fixed point of the theory as long as the one-loop coefficients obey $B_i>0$. Partially interacting fixed points are
\bea
\label{2}
&&(\al1^*,\al2^*)=\left(0,\frac{B_2}{C_2}\right)\,,\\
\label{3}
&&(\al1^*,\al2^*)=\left(\frac{B_1}{C_1},0\right)\,,
\eea
where one of the gauge coupling vanishes. The interacting fixed point is of the Banks--Zaks type \cite{Caswell:1974gg,Banks:1981nn}, provided Yukawa interactions are absent. This then also implies that the gauge coupling is asymptotically free. Alternatively, the interacting fixed point can be of the gauge-Yukawa type, provided that Yukawa couplings take an interacting fixed point themselves. In this case, and depending on the details of the Yukawa sector, the fixed point can be either IR or UV. 
Finally, we also observe fully interacting fixed points
\beq
\label{4}
(\al1^*,\al2^*)=\left(\frac{C_2B_1-B_2G_1}{C_1C_2-G_1G_2},\frac{C_1B_2-B_1G_2}{C_1C_2-G_1G_2}\right)\,.
\eeq
As such, fully interacting fixed points \eq{4} can be either UV or IR, depending on the specific field content of the theory. In all cases, we will additionally require that the couplings obey
\beq\label{pos}
\begin{array}{rcl}
\alpha_1&\ge& 0\,,\\[1ex]
\alpha_2&\ge& 0\,
\end{array}
\eeq
to ensure they reside in the physical regime of the theory \cite{Dyson:1952tj}. 

\begin{center}
\begin{table}
\begin{tabular}{cccc}
\toprule
\rowcolor{Yellow}
{}\ \ \ \bf Coupling\ \ \ &\multicolumn{3}{c}{\ \ Order in Perturbation Theory\ \ }\\
\midrule
\rowcolor{LightGray} 
$\beta_{\rm gauge}$&2&3&$n+2$ \\
\rowcolor{white}
$\beta_{\rm Yukawa}$ &1&2&$n+1$\\
\midrule
\rowcolor{LightGray}
\ \ \bf Approximation\ \ \ &{}\ \ \ \ \ LO\ \ \ \ \ \ &\ NLO\ &$n$NLO\ \\
\bottomrule
\end{tabular}
\caption{Link between approximation levels and perturbative loop orders retained in beta functions \cite{Litim:2014uca,Litim:2015iea}.}
\label{tab:tNLO}
\end{table} 
\end{center}

\subsection{Superpotential}\label{subsec:Superpotential}

In order to proceed, we must specify the superpotential/Yukawa sector. 
We assume at least two types of chiral superfields with charges under ${\cal G}_1$ and ${\cal G}_2$. At least one type of superfield must be charged under both gauge groups. Within the leading non-trivial orders in perturbation theory \cite{Bond:2016dvk,Bond:2017sem}, the beta functions for the gauge and Yukawa couplings are of the form
\beq\label{12y}
\begin{array}{rcl}
\beta_1&=&
\displaystyle
 -B_1\, \alpha_1^2+C_1\,\alpha_1^3 -D_1\,\alpha_1^2\,\al{y}+G_1\,\al1^2\,\al2\,,\\[.5ex]
\beta_2&=&
\displaystyle
 -B_2\, \al2^2+C_2\,\al2^3 -D_2\,\al2^2\,\al y +G_2\,\al2^2\,\al1\,,\\[.5ex]
\beta_y&=&
\displaystyle
 \ \ \,E\,\al{y}^2-F_1\,\al{y}\,\al 1-F_2\,\al{y}\,\al 2\,.
\end{array}
\eeq
We refer to this as the leading order (LO) approximation, see Tab.~\ref{tab:tNLO}, which is the minimal non-trivial order required to find fixed points and scaling dimensions in perturbation theory.

Fixed points of the theory are defined implicitly via the vanishing of the beta functions for all couplings. The Yukawa couplings can display either a Gaussian or an interacting fixed point
\beq\label{y}
\begin{array}{rcl}
\al y^*&=&0\,,\\
\al y^*&=&{F_1}\,\al 1^*/E\,,\\
\al y^*&=&{F_2}\,\al 2^*/E\,,\\ 
\al y^*&=&({F_1}\,\al 1^*+{F_2}\,\al 2^*)/E\,.
\end{array}
\eeq
Interacting fixed points additionally depend on whether one, the other, or both gauge couplings take an interacting fixed point alongside the Yukawa coupling. Along Yukawa nullclines \eq{y}, the system \eq{12y} reduces to \eq{beta12} whereby the two-loop coefficients $C_i$ of the gauge beta functions are shifted according to
\beq \label{C'1}
\al 2^*=0\,,\quad
\al y^*=\frac{F_1}{E}\,\al 1^*:\quad 
\left\{
\begin{array}{lcl}
C_1\to C_1'&=&
\displaystyle
C_1-D_1\,{F_1}/{E}\le C_1\,,\\[1ex]
G_2\to G_2'&=&
\displaystyle
G_2-D_2\,{F_1}/{E}\le G_2\,,\\[1ex]
B_2\to B_{2;{\rm eff}}&=&B_2
\displaystyle
-G'_2 \al 1^*\,,
\end{array}
\right.
\eeq

\beq \label{C'2}
\al 1=0\,,\quad
\al y^*=\frac{F_2}{E}\,\al 2^*:\quad 
\left\{
\begin{array}{lcl}
C_2\to C_2'&=&
\displaystyle
C_2-D_2\,{F_2}/{E}\le C_2\,,\\[1ex]
G_1\to G_1'&=&
\displaystyle
G_1-D_1\,{F_2}/{E}\le G_1\,,\\[1ex]
B_1\to B_{1;{\rm eff}}&=&B_1
\displaystyle
-G'_1 \al 2^*\,,
\end{array}
\right.
\eeq

\beq \label{C'12}
\al y^*=\frac{F_1}{E}\,\al 1^*+\frac{F_2}{E}\,\al 2^*:\quad 
\left\{
\begin{array}{rcl}
C_1\to C_1'&=&
\displaystyle
C_1-D_1\,{F_1}/{E}\le C_1\,,\\[1ex]
G_1\to G_1'&=&
\displaystyle
G_1-D_1\,{F_2}/{E}\le G_1\,,\\[1ex]
C_2\to C_2'&=&
\displaystyle
C_2-D_2\,{F_2}/{E}\le C_2\,,\\[1ex]
G_2\to G_2'&=&
\displaystyle
G_2-D_2\,{F_1}/{E}\le G_2\,.
\end{array}
\right.
\eeq

In our setting, the formal fixed points \eq{1}, \eq{2}, \eq{3} and \eq{4} have multiplicities $1, 2, 2$ and $2$, respectively, leading to seven qualitatively different fixed points, FP${}_0$ -- FP${}_6$ overall. FP${}_0$ denotes the unique Gaussian fixed point. FP${}_1$, FP${}_2$ and FP${}_3$ correspond to Banks--Zaks fixed points in either one, the other, or both gauge couplings. We refer to them as \bz{1}, \bz{2}, and \bz{12}, respectively. Similarly, FP${}_4$, FP${}_5$ and FP${}_6$ are gauge-Yukawa fixed points involving one, the other, or both gauge sectors, to which we refer as \gy{1}, \gy{2}, and \gy{12}, see Tab.~\ref{tab:tFPs}. The fixed points \bz{12} and \gy{12} are said to be fully interacting, with both gauge sectors interacting, while the fixed points \bz{1}, \bz{2}, \gy{1}, and \gy{2} are said to be partially interacting.

\begin{table*}
\begin{center}
\begin{tabular}{ccccc}
\toprule
\rowcolor{Yellow}
{\bf Fixed Point }
&
${}\quad \bm{\al 1^*}\quad$
&
${}\quad \bm{\al 2^*}\quad$
&
${}\quad \quad \bm{\al y^*}\quad \quad$
&
{\bf Type}
\\
\midrule
&
&&& \\[-2.5ex]
\BZ1
&$\displaystyle\0{B_1}{C_1}$&0&0
&\bf BZ$\, \times\, {\rm \bf G}$
\\[1.5ex]
&
&&&\\[-2.5ex]
\BZ2
&0&$\displaystyle\0{B_2}{C_2}$&0
&
\bf G{$\,\times\,$}BZ
\\[1.5ex]
&
&&& \\[-2.5ex]
\BZ{12}
&$\ \displaystyle\frac{C_2B_1-B_2G_1}{C_1C_2-G_1G_2}\ $
&$\ \displaystyle\frac{C_1B_2-B_1G_2}{C_1C_2-G_1G_2}\ $&0
&\bf BZ{$\,\times\,$}BZ
\\[1.5ex]
\midrule
&&&& \\[-2.5ex]
\GY1
&$\displaystyle\0{B_1}{{C'_1}}$&0&$\displaystyle\0{F_1}{E}\,\al 1$
&\bf \bf GY{$\,\times\,$}G
\\[1.5ex]
&&&& \\[-2.5ex]
\GY2
&0&$\displaystyle\0{B_2}{{C'_2}}$&$\displaystyle\0{F_2}{E}\,\al 2$
&\bf G{$\,\times\,$}GY
\\[1.5ex]
&&&& \\[-2.5ex]
\GY{12}
&$\displaystyle\frac{{C'_2}B_1-B_2G'_1}{{C'_1C'_2}-G'_1G'_2}$
&$\displaystyle\frac{{C'_1}B_2-B_1G'_2}{{C'_1C'_2}-G'_1G'_2}$
&$\displaystyle\frac{F_1}{E}\,\al 1+\frac{F_2}{E}\,\al 2$
&\bf GY{$\,\times\,$}GY
\\[1.5ex]
\bottomrule
\end{tabular}
\caption{Fixed points \eq{C'1}, \eq{C'2}, or \eq{C'12} in supersymmetric gauge theories with matter and gauge group ${\cal G}_1\otimes {\cal G}_2$. We also indicate how the different fixed points are interpreted as products of the Gaussian (G), Banks--Zaks (BZ), and gauge-Yukawa (GY) fixed points as seen from the individual gauge group factors (see main text).}
\label{tab:tFPs}
\end{center}
\end{table*}

In theories where none of the fermions carry gauge charges under both gauge groups, we have that $G_1=0=G_2$. In this limit, and at the present level of approximation, the gauge sectors do not communicate with each other and the ``direct product'' interpretation of the fixed points as detailed above becomes ``exact''.\footnote{For the purpose of this work, we will find it useful to refer to the ``product'' nature of interacting fixed points even in settings with $G_1,G_2\neq 0$.} Whether any of the fixed points is factually realised in a given theory crucially depends on the values of the various loop coefficients. We defer a detailed investigation of ``minimal models'' to Sec.~\ref{sec:MM}.

\subsection{Universality}\label{sec:universality}

We briefly comment on the universal behaviour and scaling exponents at the interacting fixed points of Tab.~\ref{tab:tFPs}. Scaling exponents arise as the eigenvalues $\vartheta_i$ of the stability matrix 
\beq\label{M}
M_{ij}=\partial\beta_i/\partial\alpha_j|_*
\eeq 
at fixed points. Negative or positive eigenvalues correspond to relevant or irrelevant directions, respectively. They imply that couplings approach the fixed point following a power-law behaviour in RG momentum scale, 
\beq \alpha_i(\mu)-\alpha_i^*=\sum_n c_n\, V^n_i \,\left(\frac{\mu}{\Lambda}\right)^{\vartheta_n}+{\rm subleading}\,.\eeq 
Classically, we have $\vartheta\equiv 0$. Quantum-mechanically, and at a Gaussian fixed point, eigenvalues continue to vanish and the behaviour of couplings is determined by higher-order effects. Then, couplings are either exactly marginal $\vartheta\equiv 0$, marginally relevant $\vartheta\to 0^-$, or marginally irrelevant $\vartheta\to 0^+$. In a slight abuse of language, we will from now on denote relevant and marginally relevant ones as $\vartheta\le 0$, and vice versa for irrelevant ones. 

The fixed point \textbf{G} is Gaussian in all couplings, and the scaling of couplings is either marginally relevant or marginally irrelevant. Only if $B_i>0$ can trajectories emanate from the Gaussian, meaning that it is a UV fixed point if and only if the theory is asymptotically free in both gauge couplings. Furthermore, for UV-complete trajectories, asymptotic freedom in the gauge couplings entails asymptotic freedom in the Yukawa coupling, leading, in this case, to three marginally relevant couplings with eigenvalues
\beq\label{4R}
\vartheta_1, \vartheta_2,\vartheta_3 \le 0\,.
\eeq
The fixed points \bz{1} and \bz{2} are products of a Banks--Zaks in one gauge sector with a Gaussian fixed point in the other. Since the non-zero gauge coupling at the fixed point contributes to the effective one-loop coefficient of the Gaussian gauge sector, the scaling exponents will be of the form
\beq\label{3R}
\vartheta_1, \vartheta_2\le 0< \vartheta_3\,
\eeq
for $B_\mathrm{eff}>0$, and 
\beq\label{1R}
\vartheta_1<0\le \vartheta_2,\vartheta_3\,
\eeq
for $B_\mathrm{eff}<0$. At the fixed points \gy{1} and \gy{2}, the theory is the product of a Gaussian and a gauge-Yukawa fixed point. Consequently, four possibilities arise: Provided the theory is asymptotically free, the gauge-Yukawa fixed point is IR and the eigenvalue spectrum reads \eq{1R} if the effective one-loop coefficient of the free gauge sector is $B_\mathrm{eff}>0$, and
\beq\label{0R}
0\le \vartheta_1, \vartheta_2,\vartheta_3\,,
\eeq
if $B_\eff<0$. Provided the Gaussian is a saddle, the gauge-Yukawa fixed point is either an infrared sink with scaling exponents \eq{0R}, or asymptotically safe with scaling exponents \eq{1R}.

\begin{figure}[t]
\begin{center}
\includegraphics[width=.6\linewidth]{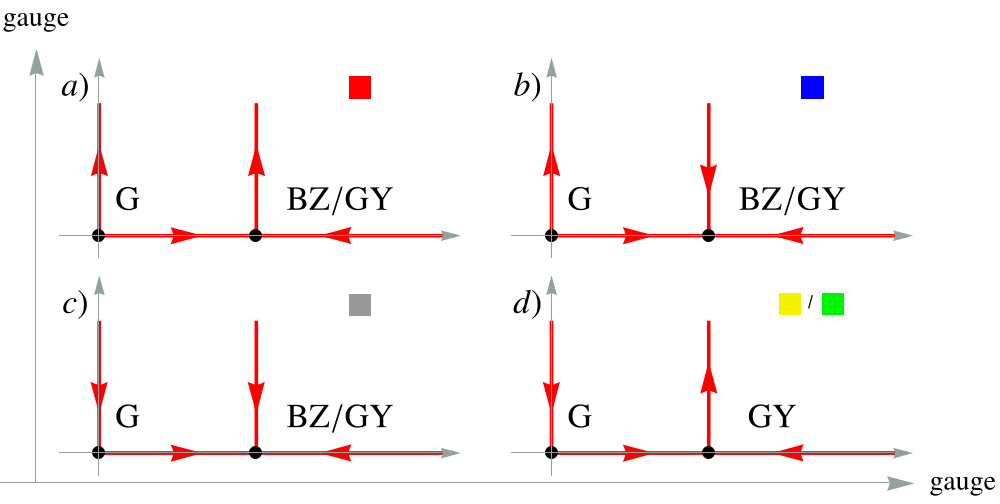}
\caption{Schematic flow diagrams for gauge couplings in the vicinity of the free (G) or interacting (BZ/GY) fixed points. If the theory is asymptotically free, the non-interacting gauge sector either remains a relevant perturbation as in panel a), or becomes irrelevant as in panel b). If the Gaussian is a saddle, the non-interacting gauge sector can either remain irrelevant as in panel c), or become relevant as in panel d). The latter is only possible for GY fixed points. If the theory is infrared free, weakly interacting fixed points are absent.}
\label{fig:relevancy_Gauge}
\end{center}
\end{figure}

In Fig.~\ref{fig:relevancy_Gauge}, we illustrate our findings for scaling exponents showing sample phase diagrams projected onto the plane of gauge couplings. Shown are the Gaussian and the Banks--Zaks or gauge-Yukawa fixed points (black dots), and RG trajectories (red) pointing from the UV to the IR. Fig.~\ref{fig:relevancy_Gauge}a) and b) relate to asymptotically free theories with either a Banks--Zaks or gauge-Yukawa fixed point in one of the gauge sectors. The difference is that the second gauge sector remains a relevant perturbation in Fig.~\ref{fig:relevancy_Gauge}a), while it becomes irrelevant in Fig.~\ref{fig:relevancy_Gauge}b) as a consequence of residual interactions. Fig.~\ref{fig:relevancy_Gauge}c) and d) relate to theories where the free fixed point is a saddle, and asymptotic freedom is absent. 
If the fixed point is Banks--Zaks or gauge-Yukawa, the second gauge sector can remain an irrelevant perturbation, see Fig.~\ref{fig:relevancy_Gauge}c). Fluctuations can also turn the marginally irrelevant gauge sector into a marginally relevant one, but {\it only} if the fixed point is of the gauge-Yukawa type, see Fig.~\ref{fig:relevancy_Gauge}d). This mechanism is key to enable asymptotically safe fixed points in what follows. We stress that Fig.~\ref{fig:relevancy_Gauge} covers all possibilities at weak coupling.

More work is required to determine the scaling exponents at the fully interacting fixed points \bz{12} and \gy{12}. To that end, we write the characteristic polynomial of the stability matrix as
\beq
\sum_{n=0}^3 T_n\,\vartheta^n=0\,.
\eeq
The coefficients $T_n$ are functions of the loop coefficients. Introducing $B=|B_1|$ and $B_2=P\,B_1$, with $P$ some free parameter, we can make a scaling analysis in the limit $B\ll 1$. Normalising the coefficient $T_3$ to $T_3=-1$, it then follows from the structure of the beta functions that $T_0={\cal O}(B^5), T_1={\cal O}(B^3)$, and $T_2={\cal O}(B)$ to leading order in $B$. In the limit where $B\ll 1$, we can deduce exact closed expressions for the leading order behaviour of the eigenvalues from solutions to the cubic equation
\beq\label{dtheta}
\begin{array}{rcl}
0&=&-\vartheta^3+ T_2 \,\vartheta^2+T_1\,\vartheta+T_0\,.
\end{array}
\eeq
The general expressions are quite lengthy and shall not be given here explicity. We note that the three eigenvalues of the three couplings at the two fully interacting fixed points \bz{12} and \gy{12} are the three solutions to \eq{dtheta} in each case. Irrespective of signs, and barring exceptional numerical cancellations, we conclude that two scaling exponents are quadratic and one is linear in $B$,
\beq\label{theta}
\begin{array}{rcl}
\vartheta_{1,2}&=&
\displaystyle
-\frac{1}{2 T_2}\left(T_1\pm\sqrt{T^2_1-4T_0\,T_2}\right)={\cal O}(B^2)\,,
\\[2ex]
\vartheta_{3}&=&T_2={\cal O}(B)\,.
\end{array}
\eeq
This is reminiscent of fixed points in gauge-Yukawa theories with a simple gauge group. The main reason for the appearance of two eigenvalues of order ${\cal O}(B^2)$ relates to the two gauge sectors, where the interacting fixed point arises through the cancellation at two-loop level. Conversely, the eigenvalue of order ${\cal O}(B)$ relates to the Yukawa coupling, as it arises from a cancellation at one-loop level. This completes the discussion of fixed points in general weakly coupled semi-simple gauge theories.

\section{\bf Minimal Models and Conformal Windows}\label{sec:MM}

In this section, we consider fixed points and conformal windows in concrete minimal models whose conformal windows are analysed to the leading non-trivial order in perturbation theory, which is two loop in the gauge and one loop in the superpotential couplings (see Tab.~\ref{tab:tNLO}).

\subsection{Semi-Simple Supersymmetric Gauge Theories}

We consider families of massless four-dimensional quantum field theories \cite{Bond:2017suy,Bond:2022xvr}
with $\Ncal=1$ supersymmetry and the semi-simple gauge group
\beq\label{gaugegroup}
\mathrm{SU}(N_1) \times \mathrm{SU}(N_2)
\eeq
keeping $N_1\ge 2$ and $N_2\ge 2$ as free parameters. We also introduce chiral superfields $(\psi,\Psi,\chi,Q)$ with multiplicities $(N_F,1,N_F,N_Q)$ and gauge charges as indicated in Tab.~\ref{tab:matter}. The superpotential to be considered is of the form
\beq\label{W}
W= y\,\Tr\big[\psi_L\,\Psi_L\,\chi_L+\psi_R\,\Psi_R\,\chi_R\big]\,,
\eeq
where the trace sums over flavour and gauge indices. Notice that the superfields $Q$ are not furnished with Yukawa interactions. Overall, the theory has a global $\mathrm{U}(N_{\rm F})_L \times \mathrm{U}(N_{\rm F})_R\times \mathrm{U}(N_{\rm Q})_L \times \mathrm{U}(N_{\rm Q})_R$ flavour symmetry and an anomaly-free $\mathrm{U}(1)_R$ symmetry. Mass terms do not affect the central conclusions and are neglected at the present stage. In four dimensions, the theory is renormalisable in perturbation theory. 

\begin{table}[t]
\begin{center}
\begin{tabular}{cc cc cc cc cc}
\toprule
\rowcolor{Yellow}
\ \ 
\bf Chiral superfields\ \ 
&$\ \bm{\psi_L}\ $ 
&$\ \bm{\psi_R}\ $ 
&$\ \bm{\Psi_L}\ $ 
&$\ \bm{\Psi_R}\ $ 
&$\ \bm{\chi_L}\ $ 
&$\ \bm{\chi_R}\ $ 
&$\ \bm{Q_L}\ $ 
&$\ \bm{Q_R}\ $ 
\\
\midrule
$\ \ \mathbf{SU}\bm{(N_1)}\ \ $
&$\overline{\Box}$
&$
\Box$
&$\Box$
&$\overline{\Box}$
&1
&1
&1
&1
\\
\rowcolor{LightGray}
\rowcolor{LightGray}
$\mathbf{SU}\bm{(N_2)}$
&1
&1
&$\Box$
&$\overline{\Box}$
&$\overline{\Box}$
&$\Box$
&$\overline{\Box}$
&$\Box$\\
\bf multiplicity
&$N_F$
&$N_F$
&$N_\Psi$
&$N_\Psi$
&$N_F$
&$N_F$
&$N_Q$
&$N_Q$
\\[.4mm]
\bottomrule
\end{tabular}
\end{center}
\vskip-.4cm
\caption{Chiral superfields and their gauge charges and flavour multiplicities.}
\label{tab:matter}
\end{table}

The theory has three classically marginal coupling constants $(g_1,g_2,y)$ given by the two usual gauge couplings and the superpotential (Yukawa) coupling, respectively. We rewrite them as
\beq
\label{couplings}
\begin{array}{lcr}
\displaystyle
\al 1=\frac{g_1^2\,N_1}{(4\pi)^2}\,,\quad
\al 2=\frac{g_2^2\,N_2}{(4\pi)^2}\,,\quad
\al y=\frac{y^{2}\,N_1}{(4\pi)^2}\,,
\end{array}
\eeq
where we have normalised the couplings with the appropriate loop factor and powers of $N_1$ and $N_2$ in view of the Veneziano limit to be adopted below.

\subsection{Free Parameters and Veneziano limit}

On the level of the Lagrangian, the free parameters of the family of theories considered are the five independent matter field multiplicities
\beq
\label{Ns} N_1, \quad N_2, \quad N_F, \quad N_\Psi, \und N_Q\,.
\eeq
\noindent
For what follows, it is convenient to parametrise the family of models using the three physically motivated parameters
\beq\label{eq:Peps}
\begin{array}{rcl}
R&=&
\displaystyle
\0{N_2}{N_1}\,,
\displaystyle
\quad P=\0{N_1}{N_2}\0{N_Q + N_F + N_\Psi N_1 - 3 N_2}{N_F + N_\Psi N_2 - 3N_1}\,, \und \eps=\0{N_F + N_\Psi N_2 - 3N_1}{N_1}\,.
\end{array}
\eeq
\noindent
$R$ denotes the ratio of the sizes of the gauge sectors. The parameter $\eps$ is the one-loop coefficient of $\beta_1$ and $P$ is the ratio of the one-loop coefficients of $\beta_2$ and $\beta_1$, up to a numerical factor. Notice that the presence of $Q$ superfields differentiates between the two gauge sectors, without which $N_1 \leftrightarrow N_2$, implying $R \leftrightarrow 1/R$, would represent the same physical theory. Observe that, instead of the five positive integers \eqref{Ns}, the parameters above \eqref{eq:Peps} (and, later, the beta functions) can be written simply in terms of the four quantities 
\begin{equation}\label{eq:ratios-of-Ns}
	\0{N_2}{N_1}\,,\quad \0{N_F}{N_1}\,,\quad \0{N_Q}{N_1}\,,\und N_\Psi\,,
\end{equation}
eliminating one degree of freedom from \eqref{Ns}. The ratios shown in \eqref{eq:ratios-of-Ns} set us up to consider the Veneziano large-$N$ limit \cite{Veneziano:1979ec}, where the field multiplicities $(N_1,N_2,N_F,N_Q)$ are sent to infinity while their ratios are kept fixed, whereby $R$, $\epsilon$, and $P$ become continuous parameters. Notice that $N_\Psi$ has to remain finite and the family of models is now parametrised by $(\eps,P,R,N_\Psi)$.

The positivity of the field multiplicities $(N_1,N_2,N_F,N_Q)$ translates to constraints in the $(\eps,P,R)$ parameters as
\beq\label{eq:positivity-constraints-full}
0<R<\frac{3+\epsilon}{N_\Psi} \und R>1+\0{(1 - R P)}{3 + N_\Psi}\eps\,.
\eeq
In a regime with strict perturbative control where
\beq\label{eps}
0 < |\eps|,|P\eps| \ll 1\,,
\eeq
the positivity constraints \eqref{eq:positivity-constraints-full} reduce to
\beq\label{eq:positivity-constraints-reduced}
1<R<\frac{3}{N_\Psi} \und P = \mathrm{finite}\,.
\eeq
The condition is non-trivial for $N_\Psi \le 3$. Ignoring the borderline cases $N_\Psi =0$ or $3$, the interesting settings relate to $N_\Psi =1$ or $2$. We consider the case $N_\Psi =1$, the reason being that it already displays the entire complexity of fixed point scenarios permitted on general grounds (Sec.~\ref{subsec:susy}), while $N_\Psi=2$ is not expected to offer qualitatively new effects. Hence, below, we employ \eqref{eq:positivity-constraints-full}, or \eq{eps} with \eqref{eq:positivity-constraints-reduced}, with $N_\Psi=1$, to understand fixed points and conformal windows.

\subsection{Banks--Zaks}\label{BZs-FPs}

\begin{center}
\begin{table}
\begin{tabular}{cllll}
\toprule
\rowcolor{Yellow}
\multicolumn{5}{c}{\ \ \bf LO beta function coefficients\ \ }\\
\midrule
\rowcolor{white}
$\beta_{1}:$ & $B_1 = -2\eps$ & $C_1 = 4(3+2\eps)$ & $D_1 = 8R(3-R\epsilon)\quad$ & $G_1 = 4R$ \\
\rowcolor{LightGray} 
$\beta_{2}:$ & $B_2 = -2 P \eps$ & $C_2 = 4(3+2 P \eps)\quad$ & $D_2 = \frac{8(3-R\epsilon)}{R}$ & $G_2 = \frac{4}{R}$ \\
\rowcolor{white}
$\beta_{\rm Yukawa}:\quad$ & $E_2 = 2(4+\eps)\quad$ & $F_1 = 4$ & $F_2 \,= 4$ &\\
\bottomrule
\end{tabular}
\caption{One- and two-loop coefficients for the gauge and Yukawa beta functions \eqref{12y}.}
\label{tab:NLO-coeffs}
\end{table} 
\end{center}

\begin{figure}[t]
\centering
\includegraphics[width=.8\linewidth]{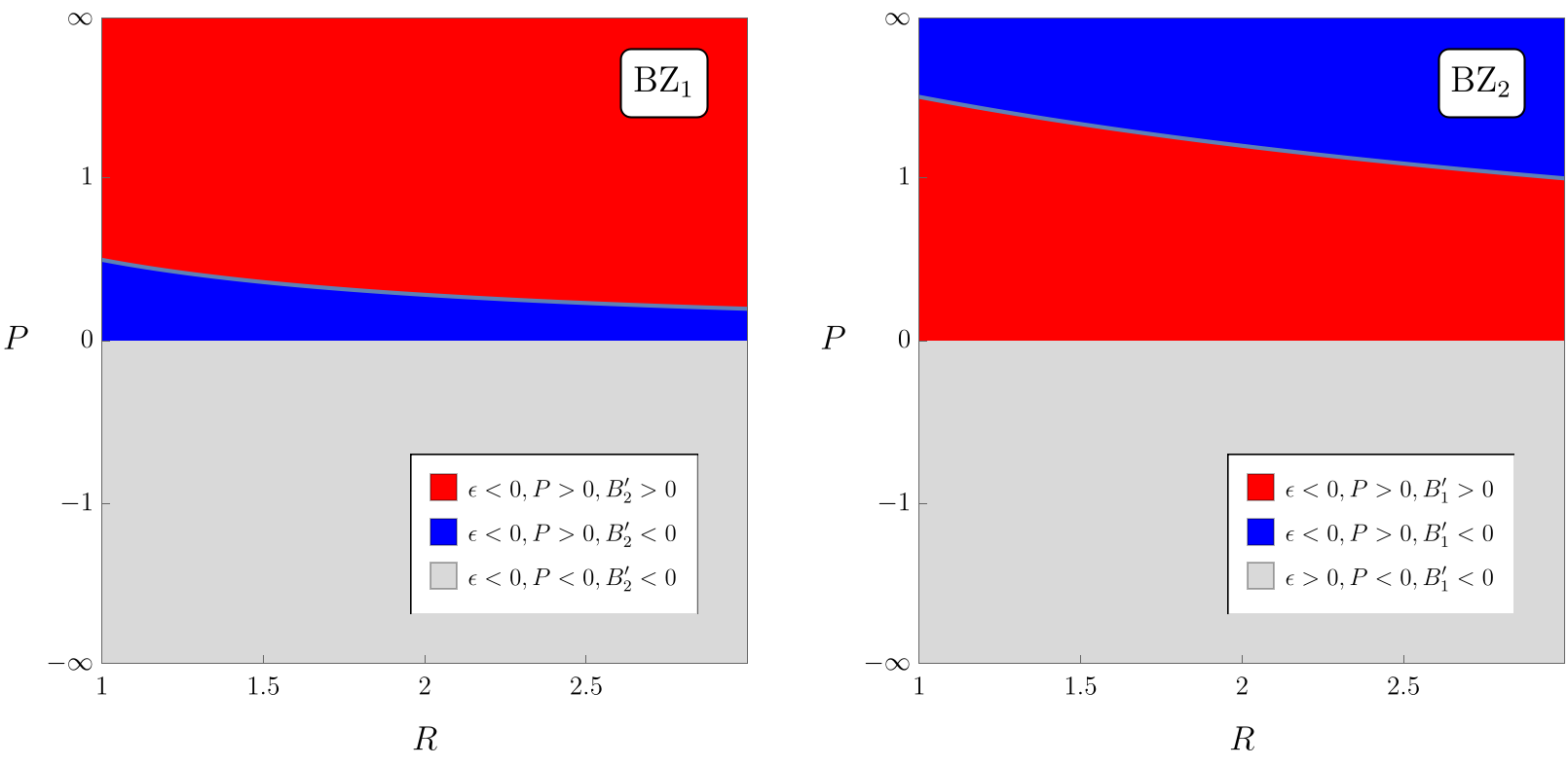}
\caption{Conformal windows of Banks--Zaks fixed points to leading order in $|\eps|\ll 1$. Note that \bz{1} and \bz{2} require $\eps<0$ and $P\eps<0$, respectively, and colours indicate the eigenvalue spectrum as in Fig.~\ref{fig:relevancy_Gauge}.}
\label{fig:CW_BZ1_BZ2_NLO}
\end{figure}

Next, we investigate the different types of fixed points one-by-one, using the beta function coefficients \eqref{12y} with loop coefficients in terms of $R$, $P$, and $\eps$ provided in Tab.~\ref{tab:NLO-coeffs}. Starting with \bz{1}, and taking the non-trivial solution of $\beta_1=0$ together with $\alpha_2^* = \alpha_y^* = 0$, we obtain
\beq\label{eq:NLO-alphas-BZ1}
\begin{aligned}
\alpha_1^*&=-\s016
\epsilon \,, \quad 
\beta_2|_*&= -B_{2;\eff}\,\alpha_2^2+ \mathcal{O}(\alpha_2^3)\,, \quad 
B_{2;\eff}= -2P\,\eps + \s0{2}{3 R}\,\eps \,.
\end{aligned}
\eeq
Here, $B_{2;\eff}$ is the effective one-loop coefficient of the second gauge sector at the non-Gaussian fixed point. Notice that, within the strictly perturbative regime of $|\eps|\ll 1$, the Banks--Zaks fixed point of the first gauge sector exists if and only if this sector is UV-free, with $\eps<0$. Moreover, the other gauge coupling will necessarily be marginal, thus, its behaviour close to the fixed point will be dictated by the sign of $B_{2;\eff}$. Its first term is the conventional one-loop coefficient, while the second one is sourced through the \bz{1}, according to \eq{C'1}. Recall that the new contribution comes from the two-loop term $G_2\alpha_1^*\alpha_2^2$ in \eqref{12y}, which is always positive, thus, $B_{2;\eff}<B_2$. Hence, residual interactions at the fixed point deflect into the non-interacting gauge sector, with the effect of making an irrelevant coupling even more irrelevant, see Fig.~\ref{fig:relevancy_Gauge}$a)$, or turning a marginally relevant coupling into a marginally irrelevant one, see Fig.~\ref{fig:relevancy_Gauge}$b)$. Here, this happens in the regime
\begin{equation}
	0<P<\s0{1}{3R}\,,
\end{equation}
as shown in the left plot of Fig.~\ref{fig:CW_BZ1_BZ2_NLO} with the different regions colour-coded as in Fig.~\ref{fig:relevancy_Gauge}.
Finally, the critical exponents of the \bz{1} fixed point are
\beq\label{eq:NLO-thetas-BZ1}
\begin{aligned}
\vartheta_1&= \s013{\epsilon^2} > 0 \,,\quad
\vartheta_2= 0 \,,\quad
\vartheta_3= \s023 \epsilon < 0 \,,
\end{aligned} 
\eeq
in line with the general findings in Sec~\ref{sec:universality}. The stability matrix is triangular, implying that $\alpha_1$ is always an irrelevant perturbation and $\alpha_y$ is a relevant one. 

Similarly, for the partially interacting fixed point \bz{2} we find
\beq\label{eq:NLO-alphas-BZ2}
\begin{aligned}
\alpha_2^*&= -\s016{P \epsilon} \,, \quad \beta_1|_*&= -B_{1;\eff}\,\alpha_1^2+ \mathcal{O}(\alpha_1^3)\,, \quad B_{1;\eff}= -2\,\eps + \s023 R\,P\,\epsilon \,.
\end{aligned}
\eeq
The fixed point is physical if and only if $\alpha_2$ is UV-free ($P\eps<0$) for any admissible $R$ and $|P|$, and the relevancy of the non-interacting $\alpha_1$ depends on the sign of $B_{1;\eff}$. The first term of $B_{1;\eff}$ is the original one-loop coefficient and the second term is sourced from interactions at the \bz{2} fixed point through the term $G_1\alpha_2^*\alpha_1^2$ term in \eqref{12y}. Once more, the original one-loop coefficient may be positive or negative, but the interaction-induced shift is always negative, making the coupling more irrelevant. This effect turns a marginally relevant coupling into a marginally irrelevant one, provided
\beq
\eps<0 \und P>\s0{3}{R}\,.
\eeq
The conformal windows is shown in the right plot of Fig.~\ref{fig:CW_BZ1_BZ2_NLO}, colour-coded according to Fig.~\ref{fig:relevancy_Gauge}. Moreover, the stability matrix is triangular, the critical exponents are
\beq\label{eq:NLO-thetas-BZ2}
\begin{aligned}
\vartheta_1&= 0^{\pm} \,,\quad
\vartheta_2= \s013{P^2\epsilon^2} > 0 \,,\quad
\vartheta_3= \s023 P\epsilon< 0 \,,
\end{aligned}
\eeq
and $\alpha_2$ and $\alpha_y$ correspond to irrelevant and relevant perturbations, respectively, while the relevancy of $\alpha_1$ depends on the sign of $B_{1;\eff}$.

\begin{figure}[t]
\includegraphics[width=0.4\linewidth]{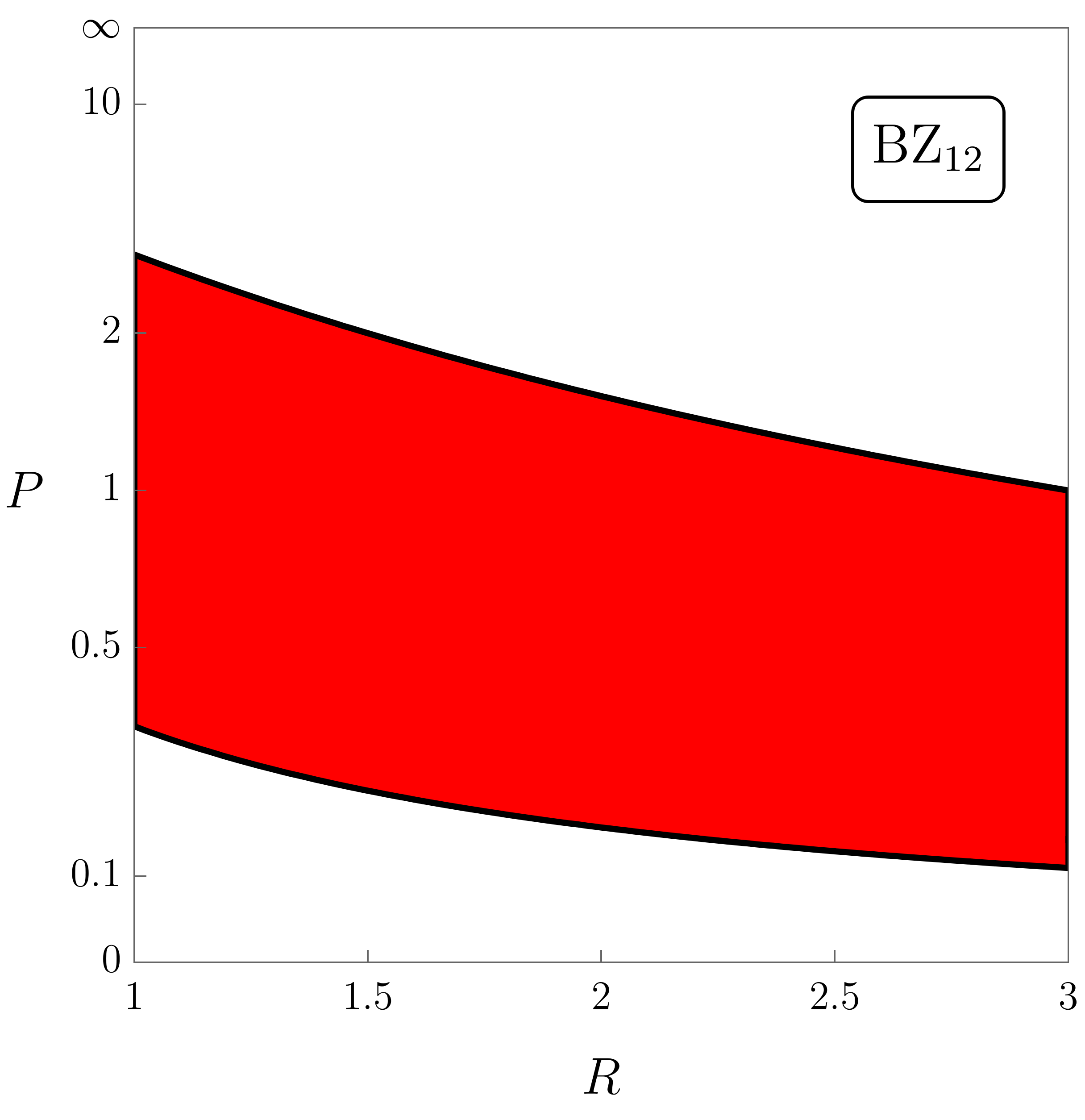}
\caption{Conformal window of the fully interacting Banks--Zaks fixed point \bz{12} for $0<-\eps\ll 1$.}
\label{fig:Win_BZ12_NLO}
\end{figure}

Finally, we look into the fully interacting Banks--Zaks fixed point, \bz{12}. Then, the values of the couplings are
\beq\label{eq:NLO-alphas-BZ12}
\begin{aligned}
\alpha_1^*&= \s01{16}{(R P - 3)}\epsilon \,, \quad 
\alpha_2^*= -\s03{16}(P-\s01{3R}) \epsilon \,, \quad 
\alpha_y^*=0\,.
\end{aligned}
\eeq
The fixed point is physical for
\begin{equation}
\eps<0 \und \s0{1}{3R}<P<\s0{3}{R}\,,
\end{equation}
which defines the parametric region illustrated in Fig.~\ref{fig:Win_BZ12_NLO}. 
Notice that the upper and lower boundaries of the \bz{12} conformal window in Fig.~\ref{fig:Win_BZ12_NLO} coincide with the $B_{2,\rm eff}=0$ and $B_{1,\rm eff}=0$ boundaries separating red- and blue-shaded areas in Fig.~\ref{fig:CW_BZ1_BZ2_NLO}, where the non-interacting gauge sector is exactly marginal. Hence, exactly at these boundaries, the fixed point \bz{12} collides either with \bz{1} or \bz{2}, leaving an exactly marginal operator in its wake. Lastly, inside the physical region of \bz{12}, $\alpha_y$ is an eigendirection of the flow. The critical exponents are
\beq\label{eq:NLO-thetas-BZ12}
\begin{aligned}
\vartheta_1 &= \0{1}{128R^2}\left(Q_1^{\mathrm{BZ}_{12}}-\sqrt{Q_2^{\mathrm{BZ}_{12}}}\right)\,\eps^2 > 0\,,\\
\vartheta_2 &= \0{1}{128R^2}\left(Q_1^{\mathrm{BZ}_{12}}+\sqrt{Q_2^{\mathrm{BZ}_{12}}}\right)\,\eps^2 > 0\,,\\
\vartheta_3 &= \left(-\0{R{-}3}{4}P + \0{3R{-}1}{4R}\right)\,\eps < 0\,,
\end{aligned}
\eeq
where $Q_1^{\mathrm{BZ}_{12}}$ and $Q_2^{\mathrm{BZ}_{12}}$ are two polynomials in $R$ and $P$ with explicit expressions given in \eqref{eq:polynomials-thetas-BZ12}. The only relevant direction is the Yukawa one. Without the superpotential, \bz{12} is the IR sink of the theory.

\subsection{Gauge-Yukawa}

\begin{figure}
\centering
\includegraphics[width=.8\linewidth]{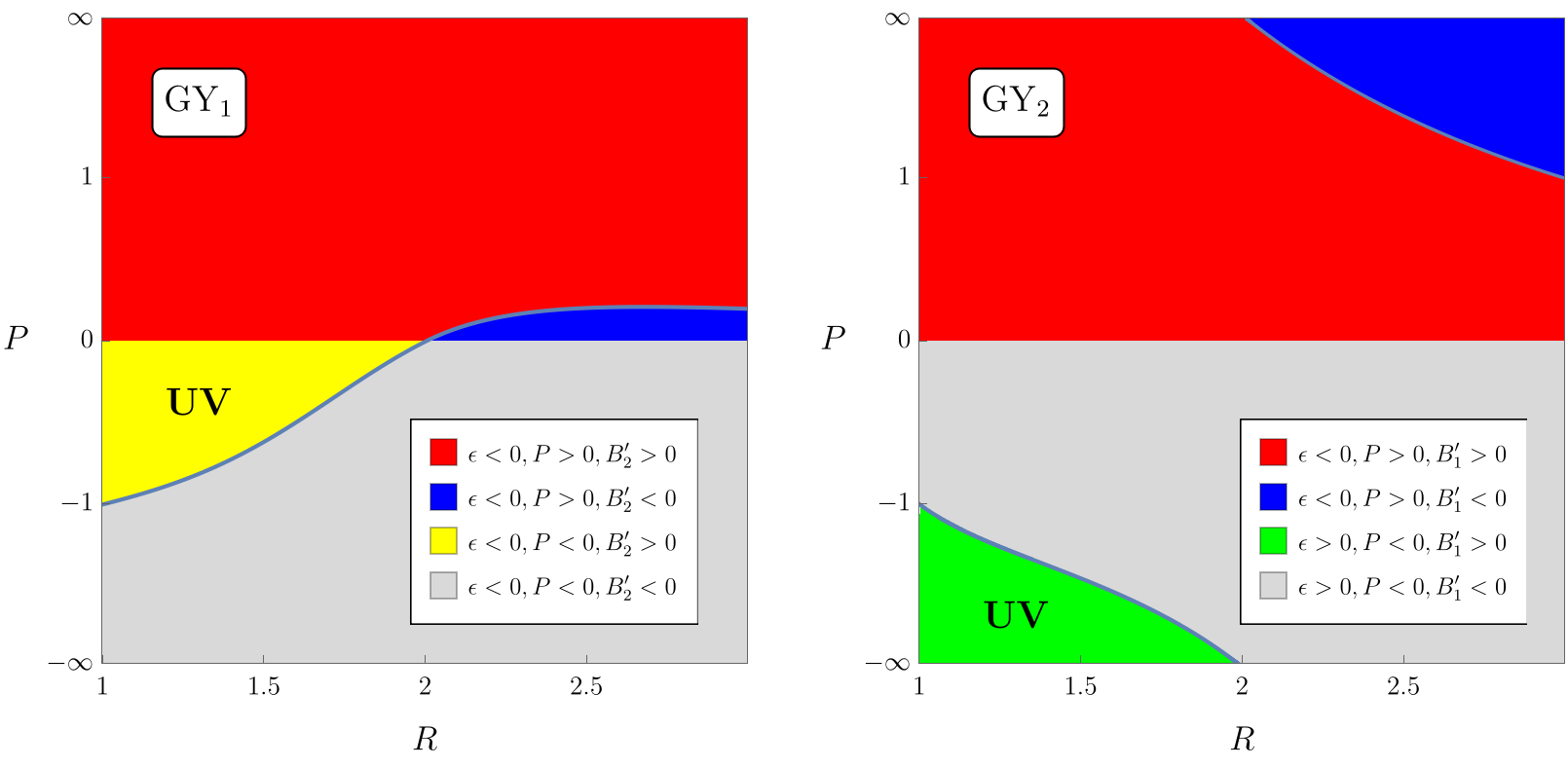}
\caption{Conformal windows of the partially interacting gauge-Yukawa fixed points as functions of $(R,P)$ and to leading order in $0<|\eps|,|P\eps|\ll 1$. Notice that \gy{1} and \gy{2} require $\eps<0$ and $P\eps<0$, respectively. The colour coding relates to the cases explained in Fig.~\ref{fig:relevancy_Gauge}. In some parameter regions (yellow, green), these fixed points are ultraviolet (UV) and asymptotically safe.}
\label{fig:CW_GY1_GY2_NLO}
\end{figure}

Next, we consider the fixed points in the Yukawa sector. Starting with \gy{1}, the couplings read
\beq\label{eq:NLO-alphas-GY1}
\begin{aligned}
\alpha_1^*&= -\frac{\epsilon}{2(R^2{-}3R{+}3)} \,, \quad
\alpha_2^*=0\,,\quad
\alpha_y^*= -\frac{\epsilon}{4 (R^2{-}3R{+}3)} \,, \\
\beta_2|_*&= -B_{2;\eff}\,\alpha_2^2 + \mathcal{O}(\alpha_2^3)\,, \quad \textrm{where}\quad
B_{2;\eff} = -2P\,\eps + \0{2(R-2)\epsilon}{R(R^2{-}3R{+}3)} \,.
\end{aligned}
\eeq
The fixed point is physical $(\alpha_1, \alpha_y>0)$ provided $\alpha_1$ is UV-free ($\eps<0$). What is new here, as opposed to Banks--Zaks type fixed points, is that the contribution to $B_{2;\eff}$ sourced by the fixed point, coming from the $(-D_1\,\al{y}+G_1\,\al{2})\,\al1^2$ terms in~\eqref{12y}, can be both negative or positive. Therefore, for certain values of the parameters, it may turn the second gauge sector from marginally irrelevant around the Gaussian to marginally relevant around the \gy{1}. Indeed, it happens whenever
\begin{equation}\label{eq:NLO-AScond-GY1}
	1<R<2\und 0 > P > -\0{2{-}R}{R(R^2{-}3R{+}3)}\,,
\end{equation}
in which the gauge sectors flow as in $d)$ of Fig.~\ref{fig:relevancy_Gauge}. This novel feature enables the \gy{1} to be a true UV fixed point, rendering the theory asymptotically safe. In Sec.~\ref{sec:PD}, Fig.~\ref{fig:Flow-region-F}, we show the corresponding UV-IR connecting trajetory of this asymptotically safe theory and, in Sec.~\ref{sec:CW}, we go beyond the next-to-leading order and we explore more in-depth the parametric region in which asymptotic safety is present. The results of \eqref{eq:NLO-alphas-GY1} are illustrated in the left plot of Fig.~\ref{fig:CW_GY1_GY2_NLO} with the appropriate colour-coding from Fig.~\ref{fig:relevancy_Gauge}. Moreover, the critical exponents read
\beq\label{eq:NLO-thetas-GY1}
\vartheta_1= \frac{\epsilon^2}{R^2{-}3R{+3}} > 0 \,, \qquad
\vartheta_2= 0\,, \qquad 
\vartheta_3= -\frac{2\epsilon}{R^2{-}3R{+3}} > 0 \,,
\eeq
with both $\alpha_1$ and $\alpha_y$ being irrelevant eigendirections of the flow for all values of $(R,P)$.

The analysis of the \gy{2} fixed point is similar, with the fixed point being physical if and only if the second gauge sector is UV-free, with $P\eps<0$. However, as asymptotic safety requires $P<0$, it may only be present for $\eps>0$ instead. The couplings at the fixed point read
\begin{align}\label{eq:NLO-alphas-GY2}
\alpha_1^*&=0\,, \quad 
\alpha_2^*= -\frac{R}{2(4R{-}3)}P\epsilon\,, \quad 
\alpha_y^*= -\frac{R}{4(4R{-}3)}P\epsilon \,, \\
\beta_1|_*&= -B_{1;\eff}\,\alpha_1^2 + \mathcal{O}(\alpha_1^3)\,, \quad \textrm{where} \quad 
B_{1;\eff} = -2\,\eps + \0{2R^2(R-2)}{4R-3}P\eps \,. \nonumber
\end{align}
Both the original one-loop coefficient $B_1$ and the new contribution to $B_{1;\eff}$ sourced from the \gy{2} may be positive or negative depending on the sign of $\eps$ and on the numerical values of $R$ and $P$, so this scenario can also reproduce the four cases in Fig.~\ref{fig:relevancy_Gauge}. The green region, in which asymptotic safety is possible, is limited by
\begin{equation}\label{eq:NLO-AScond-GY2}
	\eps>0\,,\quad 1<R<2\,,\und P < -\0{4R{-}3}{R^2(2{-}R)} < 0\,.
\end{equation}
The critical exponents for the \gy{2} are
\beq\label{eq:NLO-thetas-GY2}
\begin{aligned}
\vartheta_1&= 0 \,,\quad
\vartheta_2= \frac{R}{4R{-}3}P^2\epsilon^2 > 0 \,,\quad
\vartheta_3= -\frac{2R }{4R{-}3}P\epsilon > 0 \,, 
\end{aligned} 
\eeq
with $\alpha_2$ and $\alpha_y$ always being irrelevant eigendirections whenever the fixed point is physical, with $P\eps<0$.

\begin{figure}
\includegraphics[width=0.35\linewidth]{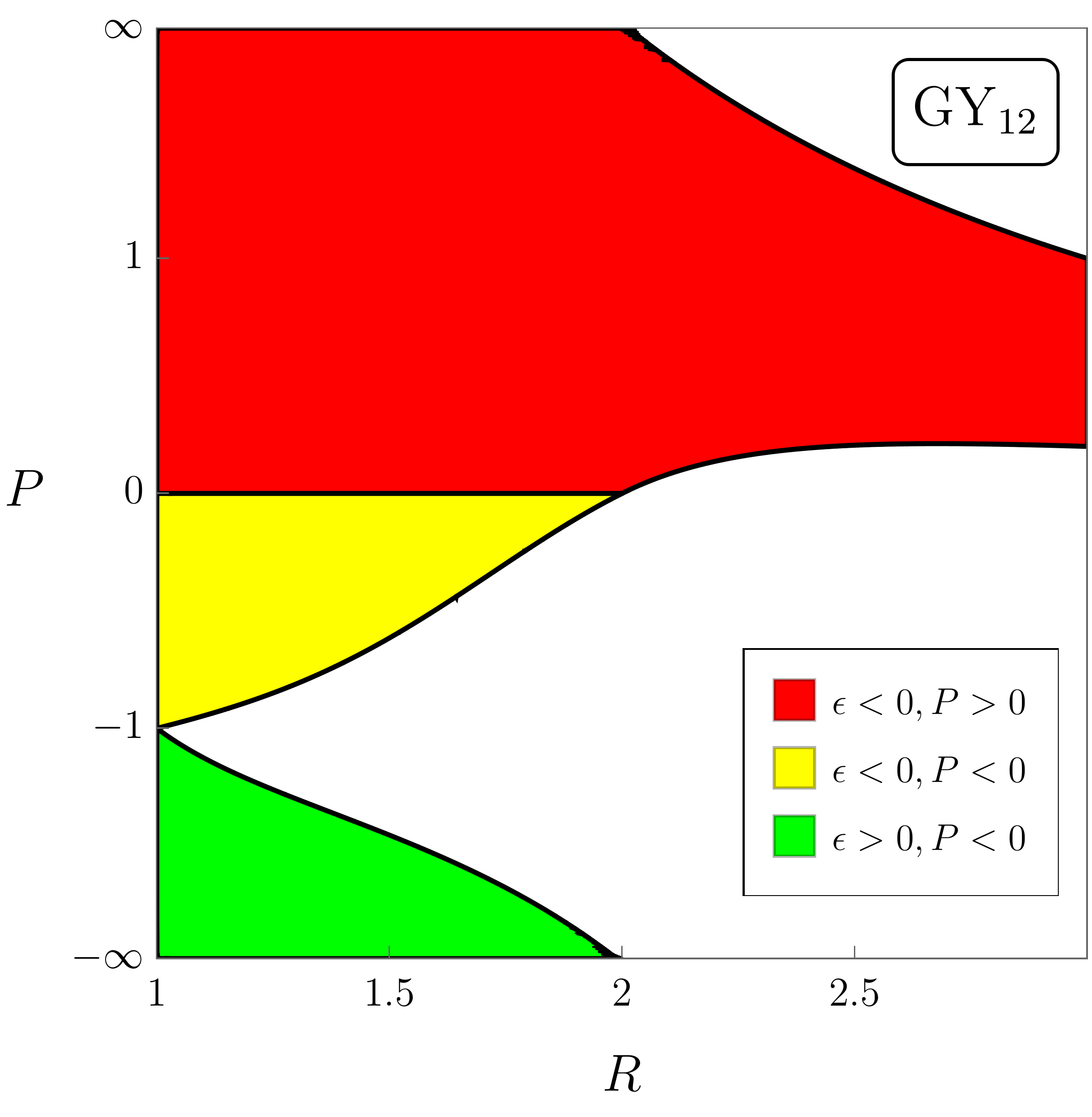}
\caption{Conformal window of the fully interacting gauge-Yukawa fixed point in the perturbative regime where $0<|\eps|\ll 1$. Whenever it exists, the fixed point \gy{12} is an infrared sink.}
\label{fig:CW_GY12_NLO}
\end{figure}

Finally, we consider the fully interacting \gy{12} fixed point. In this case, the couplings at the fixed point are
\beq\label{eq:NLO-alphas-GY12}
\begin{aligned}
\alpha_1^*&= \frac{R^2 (R {-} 2) P - (4 R {-} 3)}{2(R{-}1)(3R^2{-}8R{+}9)} \,\epsilon \,,\\
\alpha_2^*&= -\frac{R(R^2{-}3R{+}3)P-(R{-}2)}{2 (R {-} 1) (3R^2{-}8R{+}9)}\,\epsilon \,,\\
\alpha_y^*&= \frac{R(R{-}3)P-(3R{-}1)}{4(R{-}1)(3R^2{-}8R{+}9)}\,\epsilon \,,
\end{aligned}
\eeq
and the parametric regions in which the fixed point is physical are illustrated in Fig.~\ref{fig:CW_GY12_NLO}. The explicit expressions for the critical exponents of \gy{12} are too lengthy and not very enlightening, so we omit them and just point out that, whenever \gy{12} is physical, its critical exponents are all negative, so it is the fully attractive IR sink of the theory.

\section{\bf Phase Diagrams}
\label{sec:PD}

In the previous section, we have identified the different types of fixed points that can arise and their conformal windows as functions of field multiplicities. In this section, we put these findings together and study which types of fixed points are available for given field multiplicity parameters $R$ and $P$, and how the fixed points determine the phase diagrams for any given theory.

\subsection{Classification}
\begin{figure}
\includegraphics[width=0.4\linewidth]{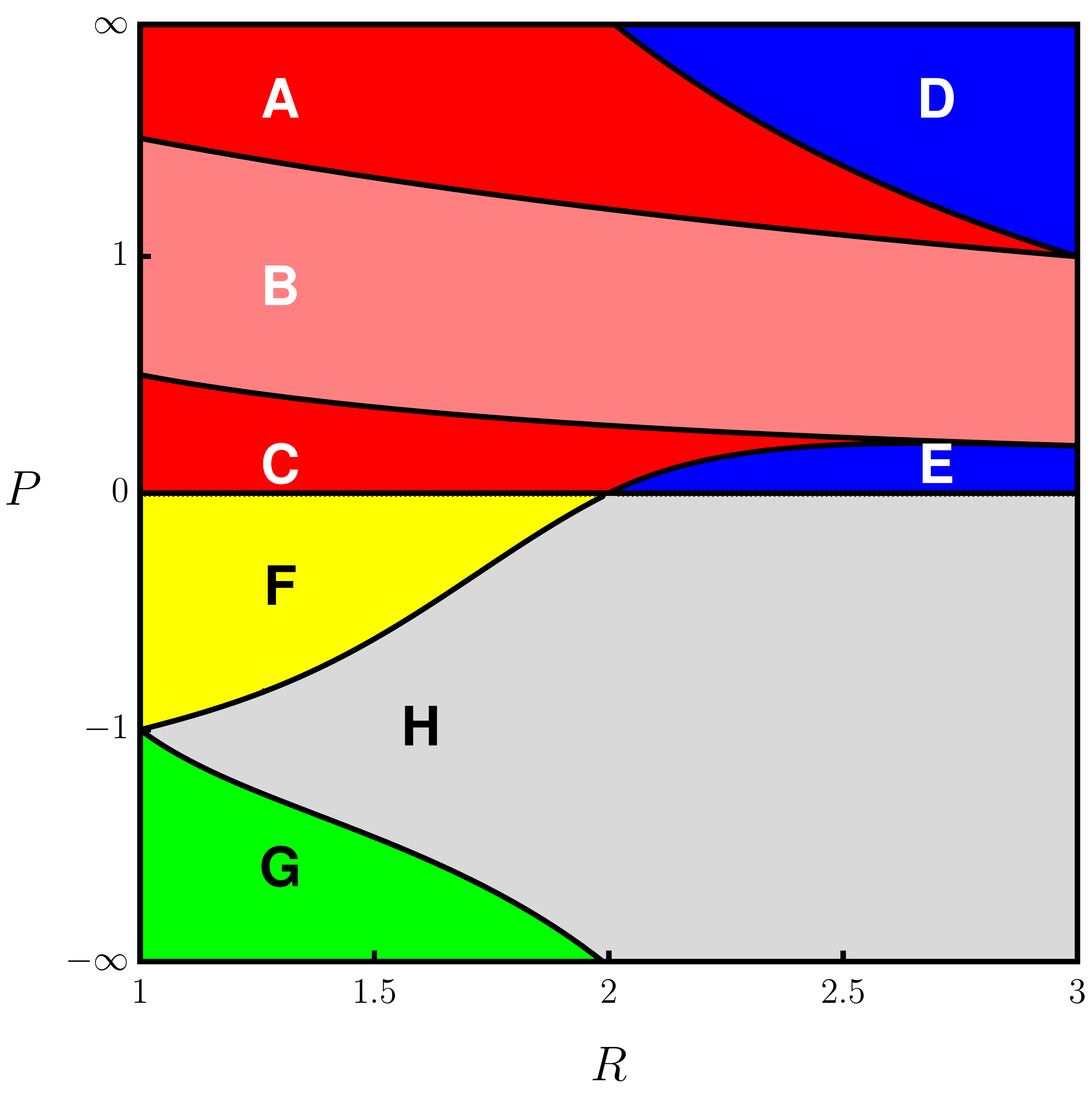}
\caption{The complete ``phase space'' of semi-simple supersymmetric quantum field theories with perturbatively controlled fixed points as functions of field multiplicities $(R,P)$, see \eq{eq:Peps}. We observe eight distinct parameter regions (A -- H), each of which is characterised by their sets of fixed points and scaling dimensions, as summarised in Tabs.~\ref{tab:pAll_AF} and~\ref{tab:pAll_AS}.}
\label{fig:pFPall}
\end{figure}

In Fig.~\ref{fig:pFPall}, we summarise results for the qualitatively different types of quantum field theories in view of their fixed points at weak coupling, together with their behaviour in the deep UV and IR. Theories differ through their matter multiplicities, which translate to the parameters $R$ and $P$, and the sign of $\eps$. Consequently, the complete ``phase space'' of qualitatively different semi-simple supersymmetric quantum field theories with perturbatively controlled fixed points shown in Fig.~\ref{fig:pFPall} arises from the overlay of the different conformal windows shown in Figs.~\ref{fig:CW_BZ1_BZ2_NLO},~\ref{fig:Win_BZ12_NLO},~\ref{fig:CW_GY1_GY2_NLO} and \ref{fig:CW_GY12_NLO}.

We observe eight distinct parameter regions A -- H, each of which is characterised by sets of fixed points and scaling dimensions. Together with the sign of $\eps$ as a free parameter, this would lead to $8\times 2=16$ different cases. However, interacting fixed points in the regions A -- E {\it only} arise for asymptotically free theories where $\eps<0$ and $P>0$, whereas fixed points in the regions F -- H can arise for either sign of $\eps$. This leaves us with $5+3\times 2=11$ different cases to consider.

We also note that the boundaries between parameter regions in Fig.~\ref{fig:pFPall} relate to the disappearance of fixed points into an unphysical domain $(\alpha^*<0)$, either due to a pole at parametrically strong coupling as for BZ fixed points, or due to a fixed point merger at weak coupling. Fixed point mergers entail Leigh-Strassler conformal manifolds with a line of fixed points \cite{Leigh:1995ep}. It then also follows that one of the universal eigenvalues changes sign across the boundary. Unitarity is automatically guaranteed since $|\eps|\ll 1$, and it does not offer bounds.

\begin{table}[t]
\begin{center}
\includegraphics[width=0.6\linewidth]{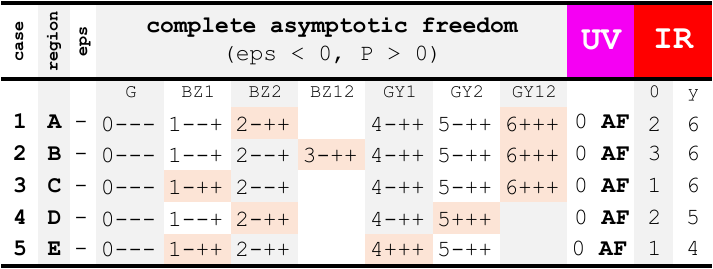}
\caption{Overview of different types of quantum field theories according to their interacting fixed points. Shown are the five distinct parameter regions of Fig.~\ref{fig:pFPall} with complete asymptotic freedom (AF). For each case, we indicate, from left to right, the corresponding parameter region in Fig.~\ref{fig:pFPall}, the sign of $\eps$, and the set of fixed points and their eigenvalue spectra (relevant: $-$, irrelevant: $+$). Orange-shaded slots highlight which fixed points are IR sinks in the absence (``0") or presence (``y") of Yukawa interactions. We observe that all possible types of fixed points are realised for any theory in the parameter region B.}
\label{tab:pAll_AF} 
\end{center}
\end{table}

\begin{figure}[b]
\begin{center}
\includegraphics[width=0.75\linewidth]{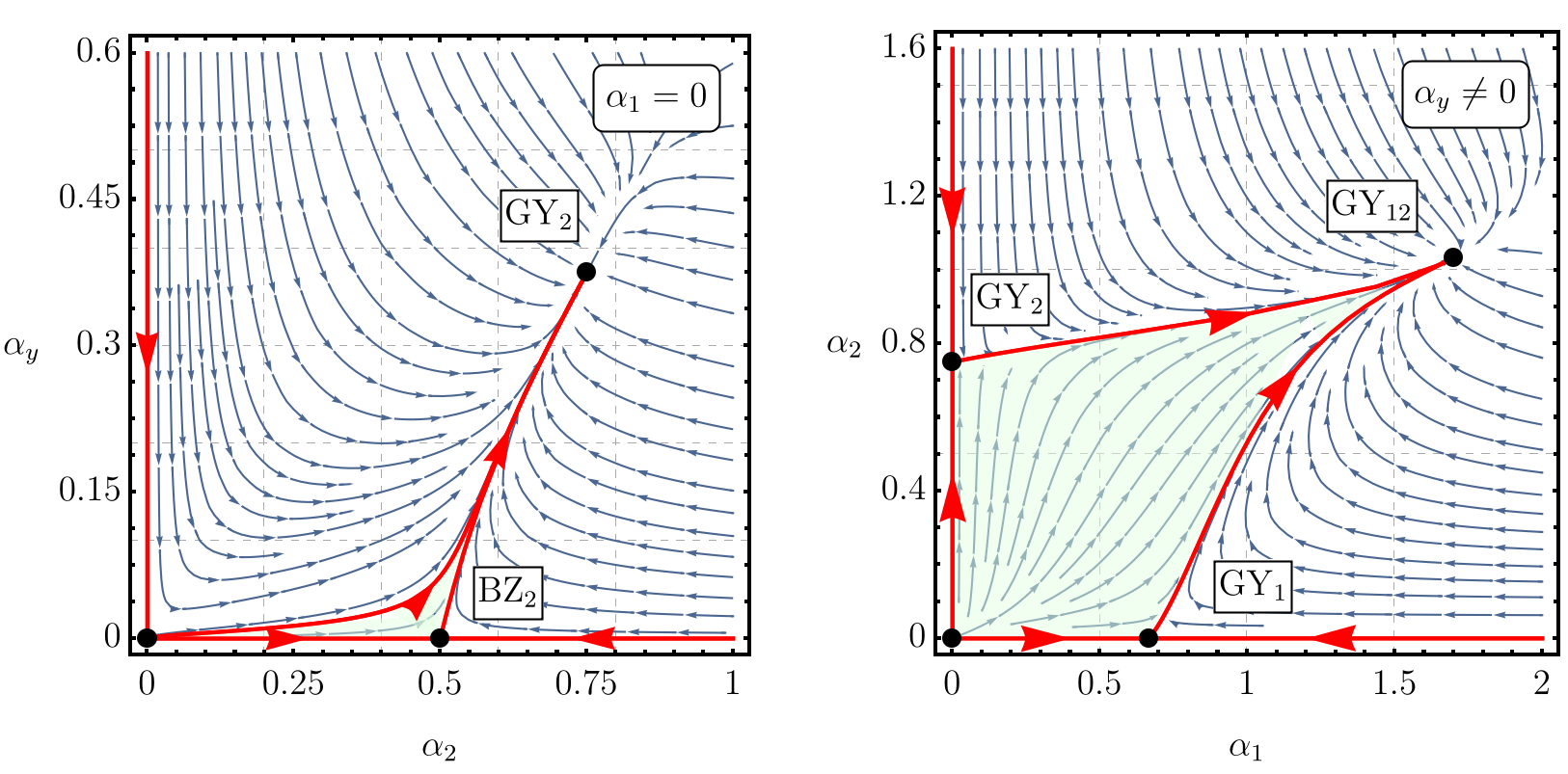}
\caption{Shown are phase diagrams for theories in the parameter region A of Fig.~\ref{tab:pAll_AF} (case 1 of Tab.~\ref{tab:pAll_AF}), projected onto the $(\alpha_2,\alpha_y)$-plane (left panel) and the ($\alpha_1,\alpha_2)$-plane (right panel). We also indicate the various fixed points (black dots), sample trajectories (blue), separatrices (red), and the set of asymptotically free trajectories (green-shaded regions) with arrows pointing towards the IR. All asymptotically free theories become conformal in the deep IR where the fixed point \gy{12} (or \bz{2} if $\alpha_y=0)$ acts as an IR attractive sink.}
\label{fig:Flow-region-A}
\end{center}
\end{figure}

\subsection{Asymptotic Freedom}

The five different cases of quantum field theories with asymptotic freedom are summarised in Tab.~\ref{tab:pAll_AF}. For each of these, the table indicates, from left to right, the corresponding parameter region in Fig.~\ref{fig:pFPall}, the sign of $\eps$, which of the seven fixed points (numbered from 0 to 6) are available, also giving their eigenvalue spectra ($-$ for each relevant and $+$ for each irrelevant eigendirection). The column “UV” indicates the UV fixed point, and the column “IR” indicates the IR fixed point depending on whether the Yukawa coupling is absent ``0'' or not ``y''. The Gaussian fixed point is always the UV fixed point, and all weakly interacting fixed points display a lower number of relevant directions, and can be reached from the Gaussian. Differences arise as to the set of physical interacting fixed points, according to Tab.~\ref{tab:pAll_AF}. Overall, theories display between four and six distinct weakly interacting fixed points. The partial Banks--Zaks fixed points (\bz{1},\bz{2}) are invariably present in all cases as a consequence of general theorems \cite{Bond:2016dvk}. We also observe that the partial gauge-Yukawa fixed points (\gy{1},\gy{2}) arise in all cases. On the other hand, the fully interacting Banks--Zaks (\bz{12}) only arises in case 2, and the fully interacting gauge-Yukawa fixed point (\gy{12}) only arises in cases 1, 2, and 3. All six distinct fixed points are available in the parameter region B (case 2).

\begin{figure}[t]
\begin{center}
\includegraphics[width=0.75\linewidth]{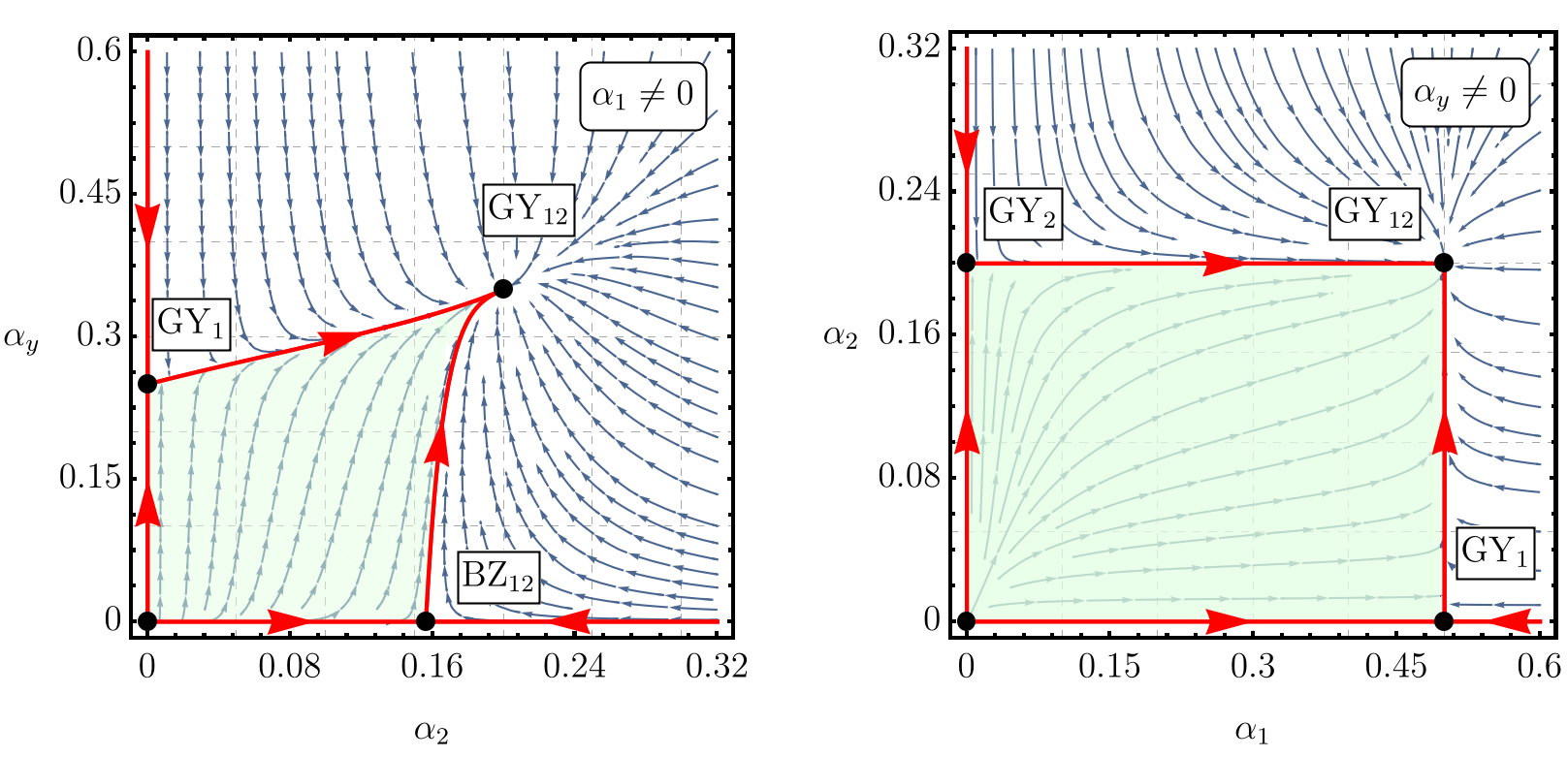}
\caption{Same as Fig.~\ref{fig:Flow-region-A} but for theories in the parameter region B of Fig.~\ref{tab:pAll_AF} (case 2 of Tab.~\ref{tab:pAll_AF}). The fixed point \gy{12} (or \bz{12} if $\alpha_y=0$) acts as an IR attractive sink for asymptotically free trajectories.}
\label{fig:Flow-region-B} 
\end{center}
\end{figure}

It is noteworthy that {\it all} theories display an interacting fixed point that acts as a fully IR attractive “sink”. In other words, all theories show a non-trivial running of couplings from the UV to the IR, yet, invariably, asymptote towards an interacting fixed point in the IR where the theory becomes superconformal. In the absence of a superpotential, the IR sink is either given by the fully interacting \bz{12} fixed point (case 2) or by one of the partially interacting \bz{1} or \bz{2} fixed points (cases 1, 3, 4, 5). With the superpotential coupling switched on, the IR sink is either given by the fully interacting \gy{12} fixed point (cases 1, 2, 3) or by one of the partially interacting \gy{1} or \gy{2} fixed points (cases 4, 5). As such, none of these asymptotically free theories can escape conformality in the deep IR. Even more so, the basin of attraction of those fixed points dominating the IR is actually larger, also attracting trajectories corresponding to UV non-complete theories, $i.e.$~trajectories not emanating from the free UV fixed point.
This pattern of results in supersymmetry is different from what has been observed in similar non-supersymmetric settings \cite{Bond:2017lnq}, where some of the asymptotically free theories escape conformality in the IR and enter regimes of strong coupling with chiral symmetry breaking and confinement.

Provided the IR sink relates to a partially interacting fixed point, it leads to a rather curious effect whereby one of the gauge sectors starts out as asymptotically free and terminates as infrared free. In other words, an asymptotically free gauge sector is turned into an infrared-free gauge sector thanks to residual conformal interactions at the IR fixed point, with trajectories that start from a vanishing gauge coupling in the UV and return to a vanishing gauge coupling in the IR, but are non-trivially interacting $\alpha>0$ in between. Examples for this behaviour are given by all theories corresponding to cases 4 and 5 of Tab.~\ref{tab:pAll_AF}, and, if $\alpha_y\equiv 0$, by cases 1 and 3.

\begin{figure}[b]
\begin{center}
\includegraphics[width=0.75\linewidth]{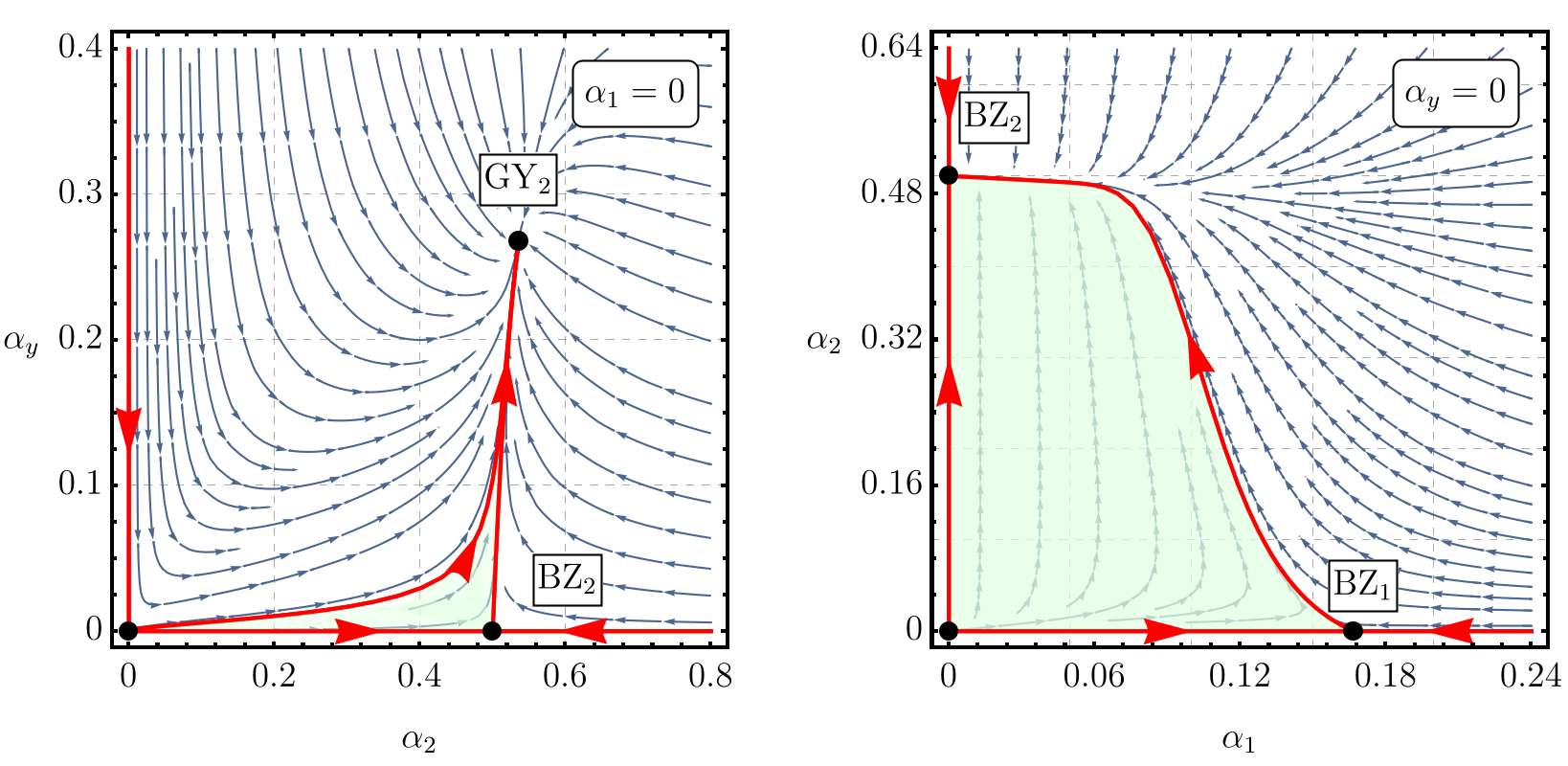}
\caption{Same as Fig.~\ref{fig:Flow-region-A} but for theories in the parameter region D of Fig.~\ref{tab:pAll_AF} (case 4 of Tab.~\ref{tab:pAll_AF}). The fixed point \gy{2} (or \bz{2} if $\alpha_y=0$) acts as an IR attractive sink for asymptotically free trajectories. Notice that the gauge sector $\alpha_1$ is both asymptotically free and infrared free, but interacting inbetween.}
\label{fig:Flow-region-D}
\end{center}
\end{figure}

Our results are further illustrated in Figs.~\ref{fig:Flow-region-A},~\ref{fig:Flow-region-B} and~\ref{fig:Flow-region-D}, where the numerical values of the couplings are shown in units of $|\eps|$. Fig.~\ref{fig:Flow-region-A} shows fixed points, phase diagrams, and trajectories representing case 1. We observe that \bz{2} and \gy{12} act as the IR sinks for all trajectories, depending on the superpotential coupling $\alpha_y$ being switched off or not. In the former case, the gauge coupling $\alpha_1$ is both UV-free and IR-free, while it remains interacting towards the IR in the latter. Fig.~\ref{fig:Flow-region-B} illustrates settings where all six types of interacting fixed points are available (case 2). Here, we observe that \gy{12} (or \bz{12} if $\alpha_y=0)$ act as IR sinks for all trajectories, which also implies that none of the gauge sectors can become free in the IR. Finally, Fig.~\ref{fig:Flow-region-D} illustrates case 4 of Tab.~\ref{tab:pAll_AF}. Here, either \bz{2} or \gy{2} act as the IR sinks for all trajectories. In either scenario, the gauge coupling $\alpha_1$ invariably becomes IR-free, courtesy of residual interactions in the deep IR at the \bz{2} and \gy{2} fixed points, respectively.

Finally, we discuss the significance of the boundaries between the regions depicted in Fig.~\ref{fig:pFPall}. The fact that they relate to fixed point mergers can now be appreciated directly from Tab.~\ref{tab:pAll_AF}. At the boundary between regions B and A, the \bz{12} fixed point merges with the \bz{2} fixed point, which is evidenced by the fact that the eigenvalue spectrum of \bz{12} is inherited by \bz{2} (see case 2 vs case 1 in Tab.~\ref{tab:pAll_AF}). After the merger, \bz{12} disappears into the unphysical domain $(\alpha^*<0)$. At the boundary, the merger generates an exactly marginal operator with a vanishing critical exponent and an associated Leigh-Strassler conformal manifold \cite{Leigh:1995ep}. Similarly, at the boundary between regions B and C, the \bz{12} fixed point merges with the \bz{1} fixed point (see case 2 vs case 3 in Tab.~\ref{tab:pAll_AF}). The boundary between regions A and D relates to a merger of the \gy{12} and the \gy{2} fixed points whereby the eigenvalue spectrum of \gy{12} is inherited by \gy{2} (see case 1 vs case 4 in Tab.~\ref{tab:pAll_AF}). After the merger, \gy{12} disappears in the unphysical domain. By the same token, we observe that the boundary between regions C and E relates to a merger of the \gy{12} and the \gy{1} fixed points, with exchange of critical exponents and the \gy{12} becoming unphysical (see case 3 vs case 5 in Tab.~\ref{tab:pAll_AF}). Either of these mergers leads to an exactly marginal operator with a vanishing scaling exponent, and an associated Leigh-Strassler conformal manifold.

\begin{table}[t]
\begin{center}
\includegraphics[width=0.6\linewidth]{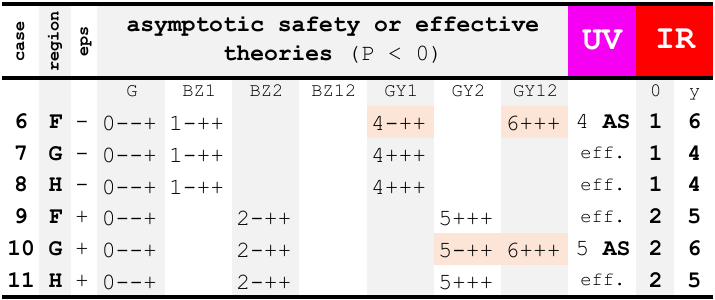}
\caption{Same as Tab.~\ref{tab:pAll_AF}, but covering the three parameter regions of Fig.~\ref{fig:pFPall} with $P<0$, characterising six distinct types of quantum field theories where the Gaussian fixed point is a saddle. Orange-shaded slots highlight asymptotically safe UV fixed points (AS) and their IR counterparts. Cases 6 and 10 represent asymptotically safe theories, while cases 7, 8, 9, and 11 correspond to effective theories.}
\label{tab:pAll_AS}
\end{center}
\end{table}

\subsection{Asymptotic Safety and Effective Theories}

Next, we discuss the six distinct cases of non-asymptotically free theories with interacting fixed points, as summarised in Tab.~\ref{tab:pAll_AS}. Here, the Gaussian fixed point is always a saddle as otherwise interacting fixed points cannot arise. For each of these, the table indicates, from left to right, the parameter region in Fig.~\ref{fig:pFPall}, the sign of $\eps$, which of the seven fixed points candidates (numbered from 0 to 6) are available, also giving their eigenvalue spectra ($-$ for each relevant and $+$ for each irrelevant eigendirection). The column “UV” indicates whether the theory is asymptotically safe (AS) or effective (eff). The column “IR” indicates the IR fixed point, depending on whether the Yukawa coupling is absent ``0'' or not ``y''.

In comparison with Tab.~\ref{tab:pAll_AF}, we observe that fixed points are more scarce. Asymptotic safety arises in two settings (cases 6 and 10). All other cases (7, 8, 9, and 11) correspond to UV incomplete theories. Partially interacting \bz{1}, \bz{2}, \gy{1} and \gy{2} are always present provided the corresponding gauge factor is UV-free \cite{Bond:2016dvk}. On the other hand, the fully interacting \bz{12} cannot arise, and the fully interacting gauge-Yukawa fixed point (\gy{12}) only arises under specific conditions such as in cases 6 and 10.

An important feature is the appearance of weakly interacting UV fixed points (cases 6 and 10) \cite{Litim:2014uca,Bond:2016dvk,Bond:2017suy}. Since the Gaussian is a saddle, it can no longer act as a UV fixed point. Its role is then taken over by the partially interacting \gy{1} (or \gy{2}), where residual interactions have turned the marginally irrelevant gauge factor $\alpha_2$ (or $\alpha_1)$ to trigger an outgoing RG flow. On the other hand, no such UV fixed point arises in cases 7, 8, 9, and 11. The reason for this is that even though the fixed points \gy{1} or \gy{2} are available, the residual interactions are not sufficient to transform the irrelevant gauge sector into a relevant one. Consequently, these renormalisable theories must be seen as effective rather than fundamental.

\begin{figure}[t]
\begin{center}
\includegraphics[width=0.38\linewidth]{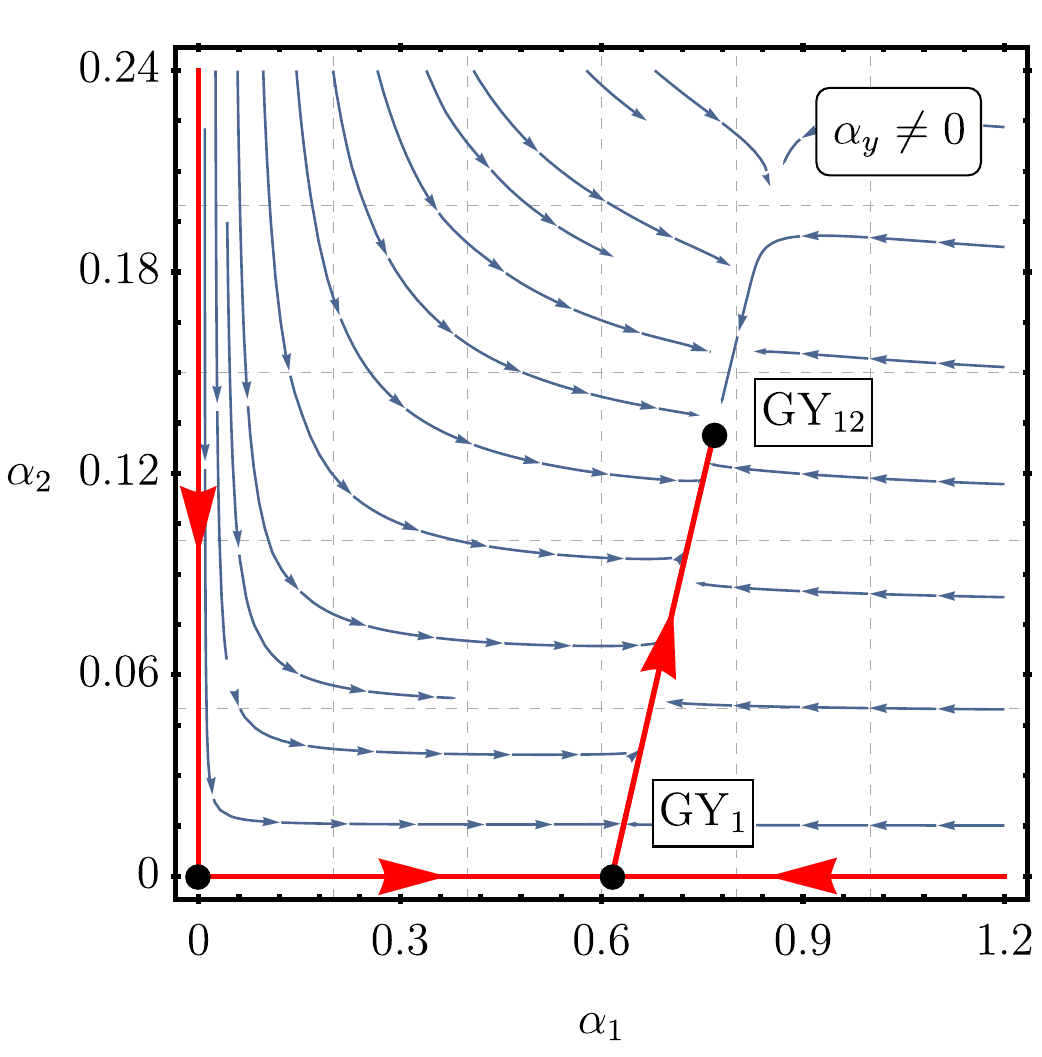}
\caption{
Phase diagram for asymptotically safe theories (case 6 of Tab.~\ref{tab:pAll_AS}), projected onto the $(\alpha_1,\alpha_2)$-plane. We indicate fixed points (black dots), sample trajectories (blue), and separatrices (red), with arrows pointing towards the IR. The interacting UV fixed point \gy{1} has a single outgoing trajectory. All theories become conformal in the deep IR where the fixed point \gy{12} acts as an IR attractive sink.}
\label{fig:Flow-region-F}
\end{center}
\end{figure}

\begin{figure}[b]
\begin{center}
\includegraphics[width=0.75\linewidth]{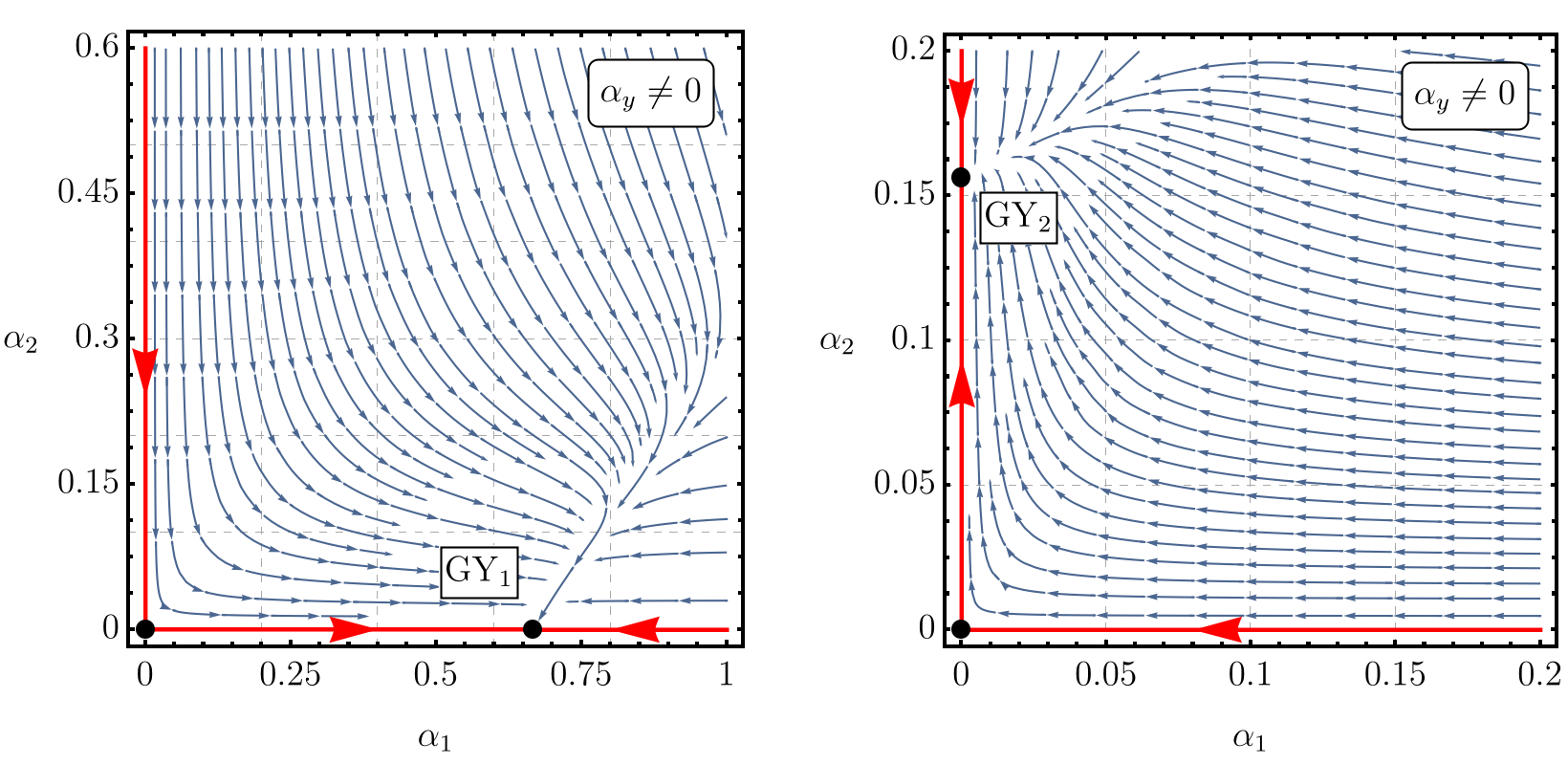}
\caption{Same as Fig.~\ref{fig:Flow-region-F}, illustrating phase diagrams in effective theories (cases 8 and 9 of Tab.~\ref{tab:pAll_AS}). All theories become conformal in the IR where the fixed point \gy{1} (left panel) or \gy{2} (right panel) acts as an IR attractor.}
\label{fig:Flow-eff}
\end{center}
\end{figure}

A curious feature of all theories in Tab.~\ref{tab:pAll_AS} is that they display interacting IR fixed points. For cases 6 and 10, these are given by the fully interacting \gy{12}. It follows that whenever a theory possesses an interacting UV fixed point, it {\it also} displays a fully interacting conformal fixed point in the IR. In all other cases, the IR sink relates to one of the partially interacting fixed points \bz{1}, \bz{2}, \gy{1}, or \gy{2}. It follows that the asymptotically non-free gauge factor is removed from the theory in the IR limit.

Finally, we discuss the boundary between regions F and G, and G and H, in Fig.~\ref{fig:pFPall}. We observe the merging of the fixed point \gy{12} with either \gy{1} or \gy{2}, evidenced by Tab.~\ref{tab:pAll_AS} where we read-off that \gy{1} inherits the eigenvalue spectrum from \gy{12} (case 6 vs case 7 or 8), and idem for \gy{2} (case 10 vs case 9 or 11). The boundary is characterised by an exactly marginal operator giving a line of IR fixed points. We conclude that asymptotic safety is available inside the regions F and G, but lost at their boundaries with H.

Our results are further illustrated in Fig.~\ref{fig:Flow-region-F} showing trajectories and the phase diagram for a scenario with an interacting UV fixed point (case 6), and in Fig.~\ref{fig:Flow-eff} for two scenarios without (cases 8 and 9). In all cases, couplings are shown in units of $|\eps|$.

\section{\bf Conformal Windows beyond Leading Orders}\label{sec:CW}

The aim of this section is twofold. Firstly, we extend the determination of fixed points and conformal scaling exponents to the next-to-leading order in $\eps$. Secondly, we exploit the higher-order corrections to estimate the 
size of conformal windows including for finite $\eps$.

The NLO study requires three-loop expressions for the gauge beta functions and two-loop results for the 
anomalous dimensions which are provided in App.~\ref{app:beta-functions-NNLO}. 
Extending the range in $\eps$ implies that bounds on the parameters $R$ and $P$ \eq{eq:positivity-constraints-reduced}, dictated by positivity of field multiplicities and assuming asymptotically small $\eps$, \eq{eps}, are now modified. Specifically, the general bound \eq{eq:positivity-constraints-full} has to be considered, which reduces to
\beq\nonumber
0<R<3+\epsilon \und R>1+\s0{\eps}{4}(1-RP)
\eeq
for $N_\Psi=1$, implying that the $R<1$ parameter region, ruled out for $|\eps|\rightarrow 0$, may become available for finite $\eps$. Similarly, parameter ranges in $P$ are equally modified depending on the sign and magnitude of $\eps$,
\beq
\begin{cases}
\eps>\operatorname{max}\big(0\,,\,4(R{-}1)\big) \geq 0 \quad &\Rightarrow \quad P>0\,,\\
0>\eps>4(R{-}1) &\Rightarrow \quad P<0\,.
\end{cases}
\eeq
Finally, we recall that unitarity dictates a bound on scalar superfield anomalous dimensions \cite{Mack:1975je}
\beq\label{eq:unitarity-bound}
\gamma_i \geq -\s012\,, \quad\mathrm{for}\quad i \in \{\psi,\Psi,\chi,Q\}\,.
\eeq
For small $|\eps|\ll 1$, we have that $|\gamma_i| \ll 1$ and constraints from unitarity are automatically satisfied. However, this can no longer be taken for granted at finite $\eps$. For the purpose of this study, when searching for zeros of beta functions, we only retain solutions that are parametrically connected with the free theory, $\alpha\to 0$ for $\eps\to 0$, and suppress (spurious) solutions that fail this criterion.

In the remainder, we focus on NLO results for the gauge-Yukawa fixed points, the reason being that these fixed points are the most relevant ones from the viewpoint of UV completing asymptotically non-free theories.
Banks--Zaks fixed points, on the other hand, even though of interest in their own right, take the role of cross-over fixed points between asymptotic UV and IR limits, whence the discussion of their conformal windows at NLO is delegated to App.~\ref{app:BZs}.

\subsection{Higher-Order Effects at \texorpdfstring{\gy{1}}{GY1}}

\begin{figure}
\includegraphics[width=0.5\linewidth]{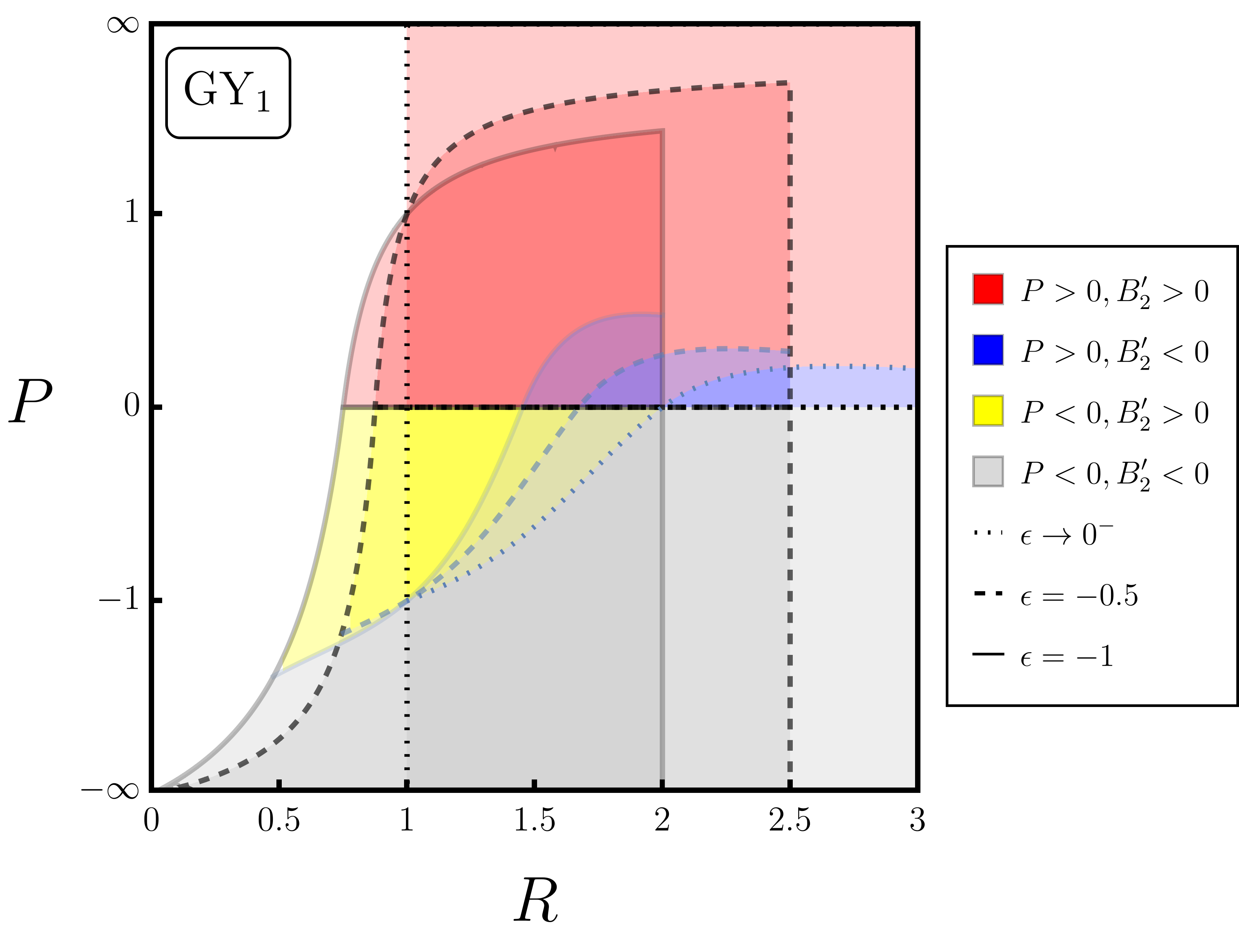}
\caption{Slices through the \gy{1} conformal window for various $\eps$.}
\label{fig:Win_GY1_NNLO}
\end{figure}

\begin{figure}[t]
\includegraphics[width=0.75\textwidth]{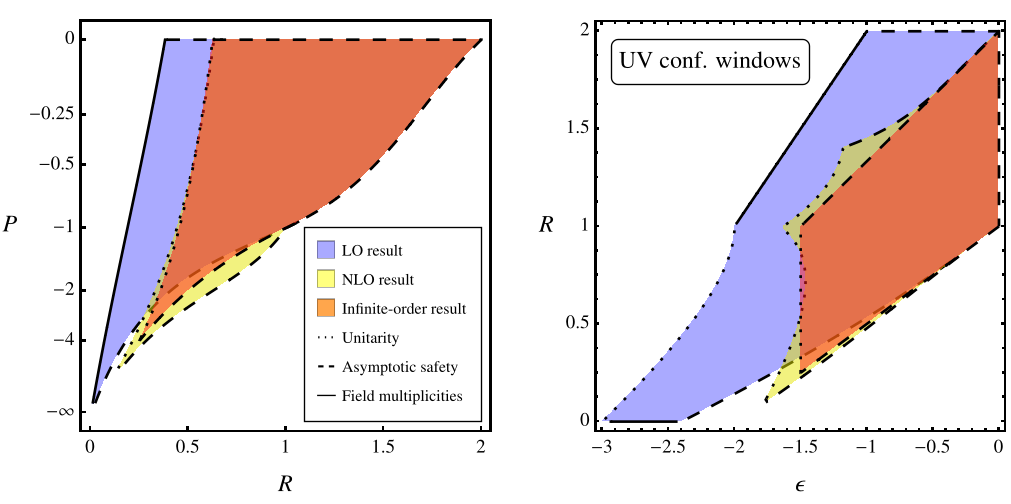}
\caption{Comparison of the \gy{1} UV conformal window at order $\eps$ (blue), order $\eps^2$ (yellow), and at infinite order (orange). It is also indicated whether boundaries arise either from unitarity (dotted line), $\alpha_2$ becoming marginally relevant (dashed), or positivity of field multiplicities (full line).}
\label{fig:GY1_projections}
\end{figure}

We begin with the gauge-Yukawa fixed point \gy{1} and recall that $\alpha_2^*=0$ corresponds to a marginal coupling. Using the beta functions up to three loop, we find the fixed point to second order in $\eps$ as
\beq\label{eq:NNLO-GY1}
\begin{aligned}
\alpha_1^{*} &= -\frac{1}{2(R^2{-}3R{+}3)}\,\eps - \frac{R^4 {+} 2 R^3 {-} 25 R^2 {+} 54 R {-} 36}{16(R^2{-}3R{+}3)^3}\,\eps^2 + \mathcal{O}(\eps^3)\,,\\
\alpha_y^{*} &= -\frac{1}{4 (R^2{-}3R{+}3)}\,\eps - \frac{(2 R - 3) (4 R^2 - 14 R + 15)}{32(R^2{-}3R{+}3)^3}\,\eps^2 + \mathcal{O}(\eps^3)\,.
\end{aligned}
\eeq
To leading order in $\eps$, positivity of couplings $\alpha_{1,y}>0$ requires $\eps<0$. At second order, however, corrections arise that may have the opposite sign. After inspection, it turns out that the subleading corrections in \eq{eq:NNLO-GY1} cannot change the sign of couplings as long as field multiplicities are within their physical domains dictated by \eq{eq:positivity-constraints-full}. We conclude that the positivity of field multiplicities is more constraining than the positivity of $\alpha^*$.
Subleading corrections also modify the effective one-loop coefficient $B_{2;\eff}$ of the marginal coupling $\alpha_2$, giving
\beq
\begin{aligned}
B_{2;\eff} &= \left(-2P + \0{2(R-2)}{R(R^2{-}3R{+}3)}\right)\,\eps + \frac{(R{-}1) (2 R^3 {-} 18 R^2 {+} 45 R {-} 39)}{4 R (R^2{-}3R{+}3)^3}\,\eps^2 + \mathcal{O}(\eps^3)\,,
\end{aligned}
\eeq
and the condition $B_{2;\eff}>0$ for $\alpha_2$ to become marginally relevant is modified accordingly,
\begin{equation}\label{eq:NNLO-AScond-GY1}
	0>P>-\frac{2-R}{R(R^2{-}3R{+}3)} + \frac{(R{-}1) (2 R^3 {-} 18 R^2 {+} 45 R {-} 39)}{8 R (R^2{-}3R{+}3)^3}\,\eps\,.
\end{equation}
We observe that the new contribution is linear in $\eps$ and its sign $\propto (R-1)\eps$, thus closing-down or opening-up parameter space in $P$ provided that $(R-1)\eps$ is positive or negative, respectively. Lastly, the critical exponents to second order in $\eps$ are found to be
\beq
\begin{aligned}
\vartheta_1 &= \frac{1}{R^2{-}3R{+3}}\eps^2 - \frac{R^2{-}7R{+}8}{4(R^2{-}3R{+3})^2}\eps^3 + \mathcal{O}(\eps^4)\,, \\
\vartheta_3 &= -\frac{2}{R^2{-}3R{+3}}\eps - \frac{4R^4{-}16R^3{+}18R^2{+}6R{-}15}{4(R^2{-}3R{+3})^3}\eps^2 + \mathcal{O}(\eps^3)\,.
\end{aligned}
\eeq
Once more, the positivity of field multiplicities together with $\eps<0$ (or $\alpha^*>0$) {\it automatically} entails that the sign of their scaling exponents is fixed to be $\vartheta_{1,3}>0$. We conclude that higher-order corrections cannot change the nature of the fixed point, even for larger $\eps$.

Recall that conformal windows are parameter ranges in $(P,R,\eps)$. To illustrate results for conformal windows at finite $\eps$, we either show projections onto two-parameter planes, or slices for fixed $\eps$. In Fig.~\ref{fig:Win_GY1_NNLO}, we consider the fixed point \eq{eq:NNLO-GY1} and show cuts through the conformal window for $\eps= 0^-,-\012,-1$. The red- and blue-shaded regions correspond to asymptotically free theories where the fixed point \eq{eq:NNLO-GY1} is IR, and where the coupling $\alpha_2$ is either marginally relevant (red) or irrelevant (blue). In the yellow-shaded region, the fixed point is UV, and $\alpha_2$ marginally relevant. We observe that the viable parameter regions shift moderately with $\eps$, and that previously inaccessible regions with $R<1$ have become available giving conformal fixed points including for larger $|\eps|$. Here, all boundaries are dictated solely by the positivity of field multiplicities.

\begin{figure}
\includegraphics[width=0.5\linewidth]{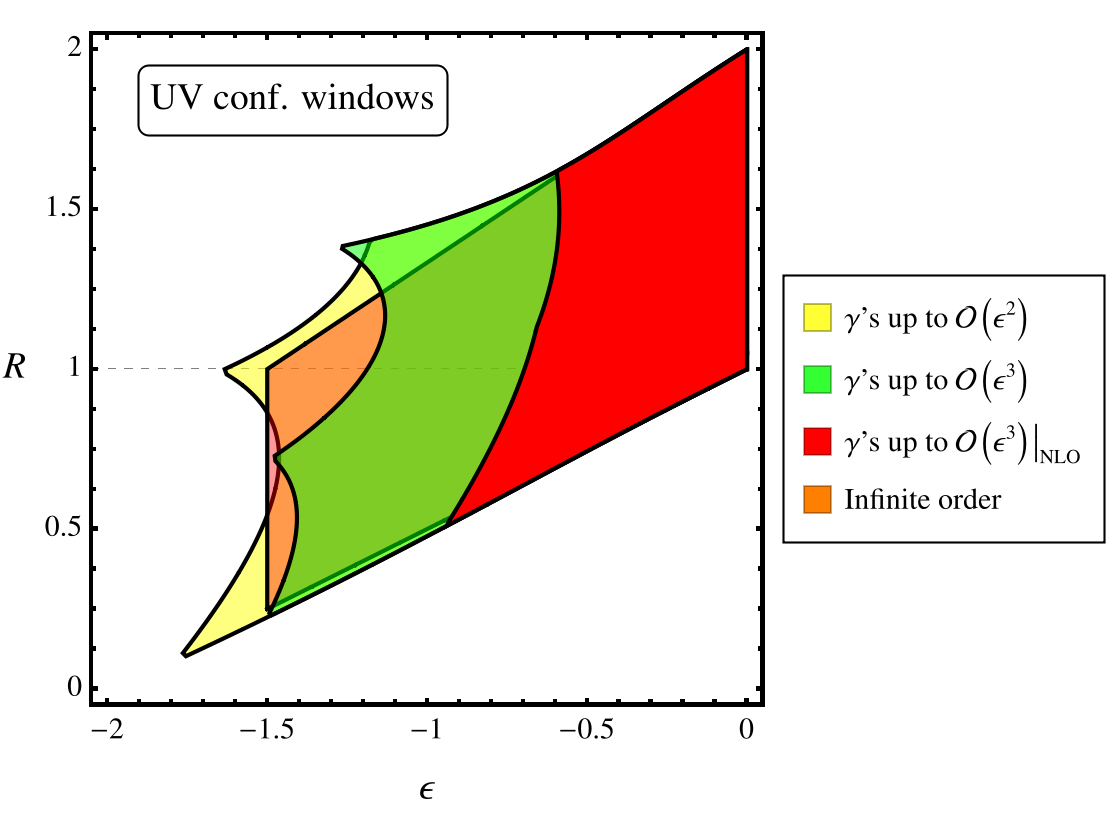}
\caption{Projection of the UV conformal window of \gy{1} onto the $(\eps,R)$-plane, comparing the NLO (yellow), NNLO (green), and the incomplete NNLO (red) approximation for the anomalous dimensions with the exact result (orange).}
\label{fig:comparison_Full_Loop_or_Eps}
\end{figure}

Let us now focus on the regime where the theory is not asymptotically free and where \gy{1} represents an UV fixed point. 
Its non-perturbative conformal window has been determined in \cite{Bond:2022xvr}. In Fig.~\ref{fig:GY1_projections}, we show projections of the UV conformal window onto the $(R,P)$ (left panel) and $(R,\eps)$ planes (right panel), also comparing the exact result (orange) with findings at LO (blue) and NLO (yellow). It is also indicated whether boundaries arise either from unitarity (dotted line), $\alpha_2$ becoming marginally relevant ($B_{2,\rm eff}>0$, dashed), or positivity of field multiplicities (full line). As expected, we find that unitarity only plays a role for higher values of $|\eps|$ such as in the $R<1$ regime. Also, unitarity constraints in Fig.~\ref{fig:GY1_projections} are mostly set by the chiral superfield $\psi$, and occasionally by $\Psi$, while the anomalous dimension of $\chi$ never violates the unitarity condition at this fixed point. We observe that the NLO results are much closer to the exact ones, and clearly improve on the findings at LO. 

\begin{figure}
\includegraphics[width=0.5\linewidth]{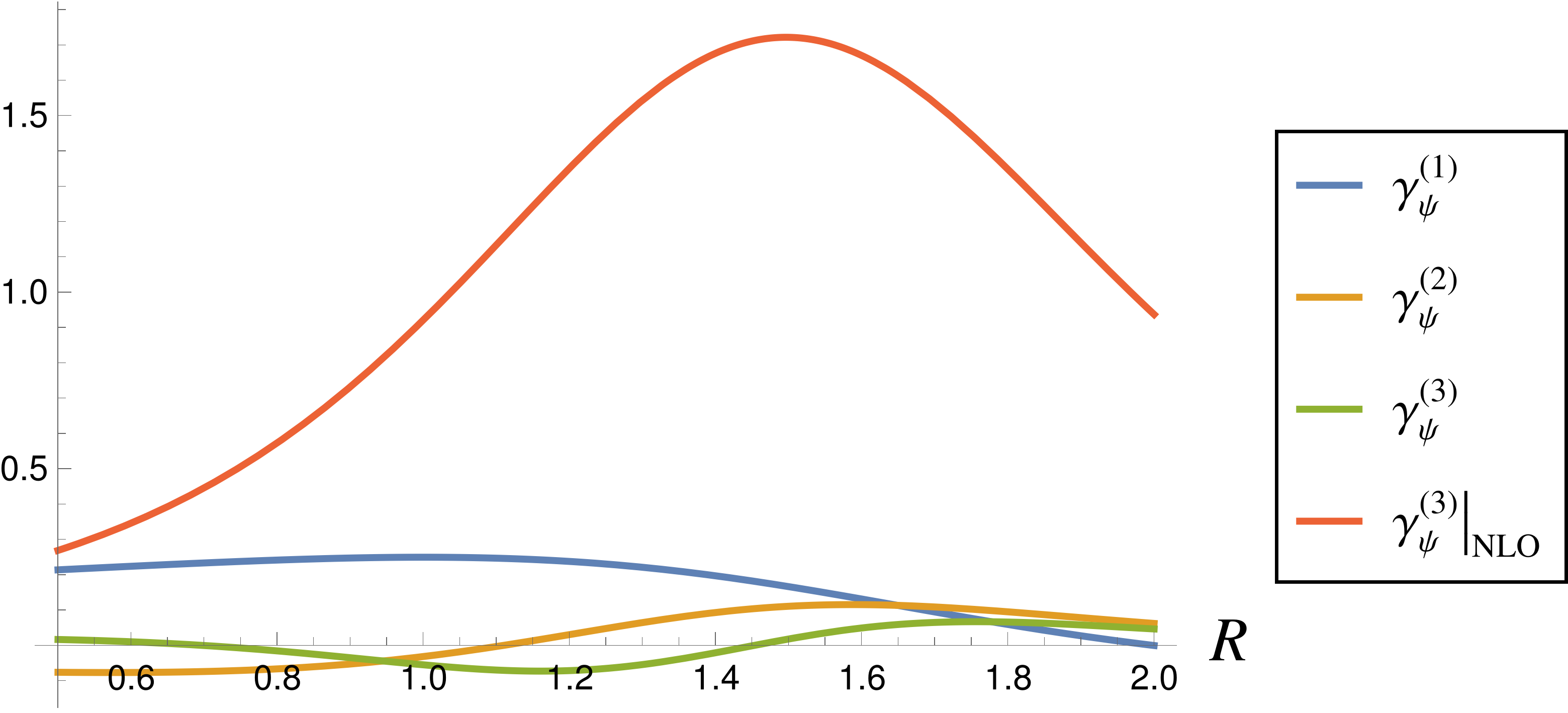}
\caption{Shown are the coefficients $\gamma_{\psi,1}$ (blue), $\gamma_{\psi,2}$ (yellow), $\gamma_{\psi,3}$ (green) and $\gamma_{\psi,3}|_{{}_{\rm NLO}}$ (orange) defined via \eq{gamma_expansion}.
Notice that $\gamma_{\psi,3}$ and $\gamma_{\psi,3}|_{{}_{\rm NLO}}$ differ significantly.}
\label{fig:comparison_gamma_3}
\end{figure}

We briefly discuss the chiral superfield anomalous dimensions in more detail. Fig.~\ref{fig:comparison_Full_Loop_or_Eps} shows the \gy{1} conformal window obtained by exploiting expressions for anomalous dimensions to NLO (yellow), NNLO (green), an {\it incomplete} NNLO (red), and the exact result (orange). ``Incomplete NNLO" refers to the expression for $\gamma$ at NLO where, in addition, those terms up to ${\cal O}(\eps^3)$ are retained that already arise at the present loop level (see Tab.~\ref{tab:tNLO}).\footnote{The approximation is incomplete because the four-loop gauge and three-loop Yukawa contributions, required to find the complete NNLO expression $\gamma^{(3)}$, are absent.} We observe that NLO and NNLO largely agree with the exact result, except close to the unitarity boundary which is most sensitive to approximations, and that the incomplete NNLO approximation is worse than the NLO and NNLO ones. The reason for this discrepancy can be understood from Fig.~\ref{fig:comparison_gamma_3}, which compares the magnitude of the expansion coefficients in
\beq\label{gamma_expansion}
\gamma=\sum_{i=1}\gamma_{i}\,\eps^i\,.
\eeq
The first three coefficients for $\gamma_{\psi,i}$ are of the same order of magnitude over the entire range of $R$. On the other hand, the incomplete coefficient $\gamma_{\psi,3}|_{{}_{\rm NLO}}$ comes out significantly larger than the exact coefficient $\gamma_{\psi,3}$.\footnote{Explicit expressions for either of these are given in the Appendix, see \eqref{eq:comparing-gamma-3-gy1} and \eqref{eq:incomplete}.} Moreover, for a significant range in $R$, $\gamma_{\psi,3}|_{{}_{\rm NLO}}$ also has the opposite sign with respect to the exact one, thereby overconstraining the unitarity bound \eq{eq:unitarity-bound} on the conformal window. We conclude that these differences are at the root of the discrepancy in Fig.~\ref{fig:comparison_Full_Loop_or_Eps}. For the purpose of determining conformal windows for larger $\eps$, {incomplete} approximations for anomalous dimensions should better be avoided.

\subsection{Higher-Order Effects at \texorpdfstring{\gy{2}}{GY2}} 

Next, we consider \gy{2}. Writing fixed points and scaling exponents as formal power series in $\eps$, we determine the next-to-leading order correction terms from the non-trivial zeros of the beta functions. For the fixed point and the effective one-loop coefficient we find
\begin{eqnarray}\label{eq:NNLO-GY2}
\nonumber
\alpha_2^*&=&-\frac{RP\epsilon }{2(4R{-}3)} + \frac{R P}{8(4R{-}3)^2} \left( \frac{R(R^3{-}3R^2{+}49R{-}39)}{2(4R{-}3)}P - (R{+}1) \right)\eps^2\\ \label{eq:GY2_NNLO_alphas}
\alpha_y^{*}&=& -\frac{RP\epsilon }{4(4R{-}3)} - \frac{R P}{16(4R-3)^2} \left( \frac{R(3R^3{-}4R^2{-}42R{+}36)}{2(4R{-}3)}P - (3R{-}4) \right)\eps^2\,, \\ \nonumber
B_{1;\eff} &=& {-}2\left(1 {-} \0{R^2(R{-}2)}{4R{-}3}P\right)\eps {+} {R^2 P}{}\left( \frac{RP(R{-}2)(R{+}3)(3R^2{-}17R{+}12)}{4(4R{-}3)^3} {-} \frac{(R{+}1)(3R{-}1)}{2(4R{-}3)^2} \right) \eps^2\ \ \ \ \ {}
\end{eqnarray}
up to higher loop corrections. Similarly, recalling that two of the three scaling exponents start out as $\sim \eps^2$ to leading order, and writing them as $\vartheta_{1,2} = \vartheta_{1,2}^{(2)}\,\eps^2 + \vartheta_{1,2}^{(3)}\,\eps^3 + \cdots$ and $\vartheta_{3} = \vartheta_{3}^{(1)}\,\eps + \vartheta_{3}^{(2)}\,\eps^2 + \cdots$, respectively, we find their subleading corrections as
\beq
\begin{aligned}
\vartheta_2^{(3)}&= - \frac{R P^2}{4(4R{-}3)^2}\big[2(5R{-}3)P-(R{+}1)\big] \,,\\
\vartheta_3^{(2)}&= \frac{R P}{4(4R{-}3)^3}\big[ (5R^4{-}18R^3{+}14R^2{+}42R{-}36)\,P - 2(R{+}1)(4R{-}3) \big]\,,
\end{aligned} 
\eeq
while the leading-order coefficients can be extracted from \eq{eq:NLO-thetas-GY2}.

\begin{figure}[t]
\includegraphics[width=0.75\textwidth]{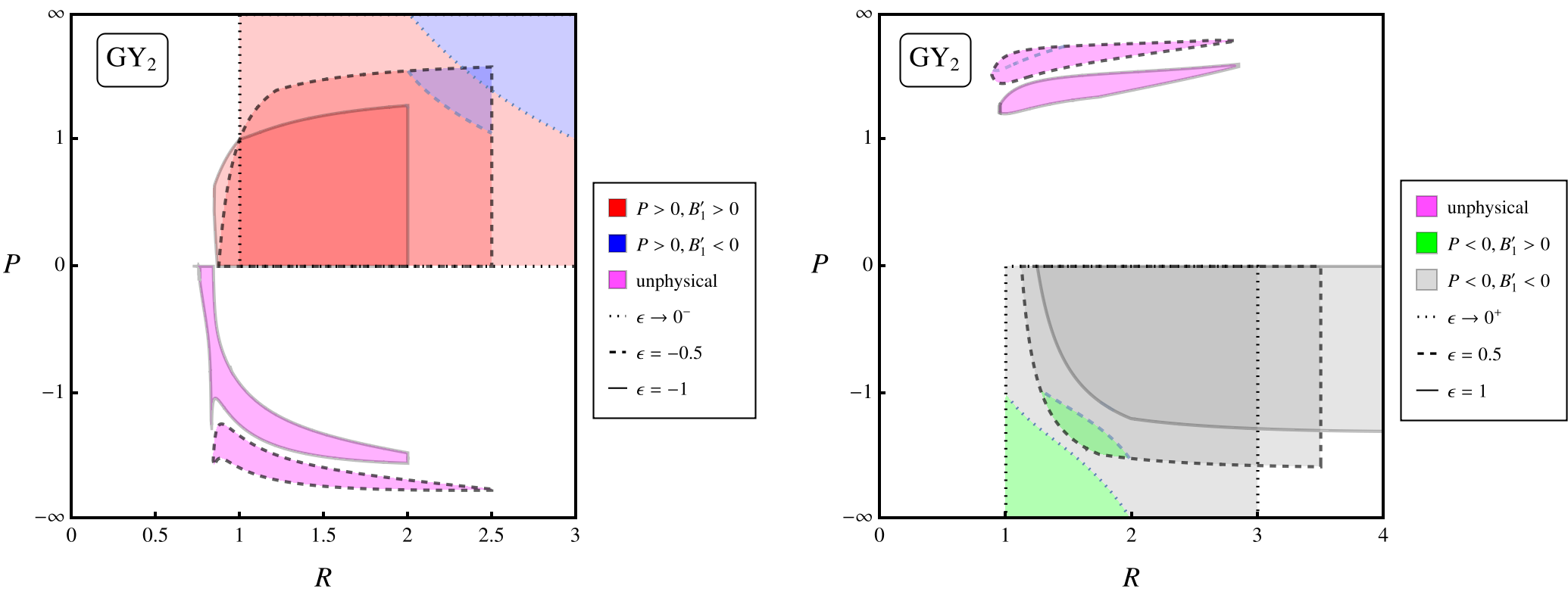}
\caption{Slices through \gy{2} conformal windows at next-to-leading order for several $\eps<0$ (left) and $\eps>0$ (right panel).}
\label{fig:GY2_NNLO_epsNeg}
\end{figure}

{The \gy{2} conformal windows at next-to-leading order are illustrated in Fig.~\ref{fig:GY2_NNLO_epsNeg}. For $\eps<0$ (left panel) and for $P>0$, viable parameter regions arise as smooth and shrinking deformations of the (red- and blue-shaded) perturbative regions $\eps\rightarrow 0^-$ (see Fig.~\ref{fig:CW_GY1_GY2_NLO}, right panel), also giving access to regions with $R<1$. For $P<0$, new regions open up (magenta) that are not present for $\eps\rightarrow 0^-$. However, these solutions are unphysical and violate the $a$-theorem, which can be seen as follows. First, notice that $P\eps>0$ implies that the interacting gauge sector is IR-free. Consequently, the fixed point itself would be UV, and in its vicinity the phase diagram would look like Fig.~\ref{fig:unphysical_relevancy}e) or f), and with RG trajectories connecting the interacting fixed point with the Gaussian. For this scenario to be compatible with the weak form of the $a$-theorem, the difference between the central charge $a$ at the interacting UV and the free IR fixed point must be positive
\begin{equation}\label{eq:constraint-a-theorem}
	a_{\mathrm{UV}}-a_{\mathrm{free}}>0 \,.
\end{equation}
Following \cite{Martin:2000cr} and using general expressions for the central charges, one finds 
\begin{align}\label{eq:adiffGauss}
	a_{\mathrm{UV}} - a_{\rm free} = \frac{9}{32} \sum_i N_i\left(R_i - \frac{2}{3}\right)^2\left( R_i - \frac{5}{3}\right)\,.
\end{align}
where $N_i$ are the number of chiral superfields with $R$-charge $R_i$ at the UV fixed point. We observe that for \eqref{eq:constraint-a-theorem} to hold true, at least one of the $R$-charges must be parametrically large, $R_i>\s053$, implying superfield anomalous dimensions $\gamma^{\mathrm{UV}}_i > \s032$ \cite{Martin:2000cr}. However, we have checked by direct inspection that this is not the case, and hence these solutions are in violation of \eq{eq:constraint-a-theorem}.

Similarly, for $\eps>0$ (Fig.~\ref{fig:GY2_NNLO_epsNeg}, right panel) viable parameter regions for larger $\eps$ arise as smooth deformations of the (green-shaded) perturbative regions where $\eps\rightarrow 0^+$ (see Fig.~\ref{fig:CW_GY1_GY2_NLO}). The region shrinks with growing $\eps$ and eventually disappears around $\eps_{\mathrm{max}}^{\mathrm{GY}_2}\approx 1.0066$. A new (magenta) region opens up once $P>0$, and where the fixed point would be UV with phase diagrams as in Fig.~\ref{fig:unphysical_relevancy}. By direct inspection, we observe anomalous dimensions within $-\012\lesssim\gamma_i\lesssim 1$, too small to satisfy the $a$-theorem \eq{eq:constraint-a-theorem}. We conclude that all fixed points in the magenta-shaded regions in Fig.~\ref{fig:GY2_NNLO_epsNeg} are unphysical and must be dropped, confirming that $\alpha_2$ must be UV-free ($P\eps<0$) in the physical region.

\begin{figure}[b]
\includegraphics[width=0.6\textwidth]{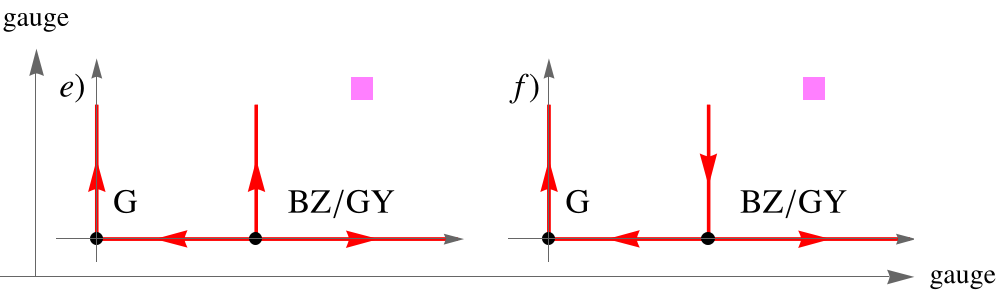}
\caption{The ``would-be" flow diagrams for fixed points in the magenta-shaded areas of Fig.~\ref{fig:GY2_NNLO_epsNeg}.
Note that these settings cannot be realised at weak coupling, much unlike those shown in Fig.~\ref{fig:relevancy_Gauge}.}
\label{fig:unphysical_relevancy}
\end{figure}

Finally, Fig.~\ref{fig:GY2_projections} illustrates how the \gy{2} conformal window grows from leading to next-to-leading order in the approximation. Shown are projections of the UV conformal window onto the $(R,P)$ and $(\eps,R)$ planes, comparing LO (blue) and NLO (yellow) results. In either case, we observe that the NLO corrections have enabled a wider parameter space. This indicates that, as soon as $\eps$ is no longer perturbatively small, a larger set of asymptotically safe quantum field theories becomes available than naively expected from perturbation theory. This is in accord with \cite{Bond:2022xvr}, which demonstrated that the \gy{1} conformal window is significantly larger than its perturbatively accessible part.

\begin{figure}[htb]
\includegraphics[width=0.75\textwidth]{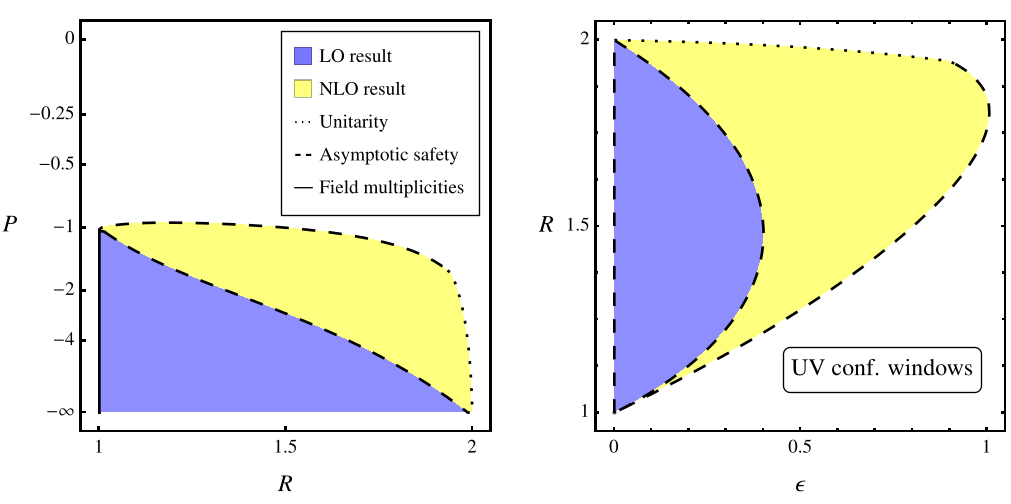}
\caption{Projections of the UV conformal window for \gy{2} at the leading (blue) and next-to-leading (yellow) order, also indicating whether boundaries are dictated by positivity of field multiplicities, $B_{\rm eff}>0$, or unitarity.}
\label{fig:GY2_projections}
\end{figure}

\subsection{Higher-Order Effects at \texorpdfstring{\gy{12}}{GY12}}

\begin{figure}[htb]
\includegraphics[width=0.75\textwidth]{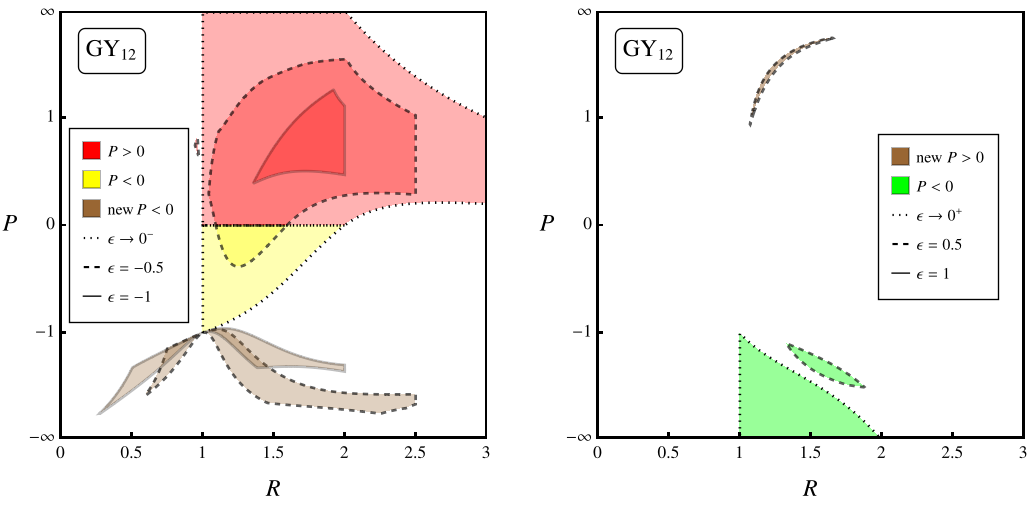}
\caption{Slices of the \gy{12} conformal window at next-to-leading order covering $\eps<0$ (left panel) and $\eps>0$ (right).}
\label{fig:Win_GY12_NNLO}
\end{figure}

Lastly, we look into the fixed point \gy{12}. At the next-to-leading order, we find
\beq\label{eq:alphas-GY12-NNLO}
\begin{aligned}
\alpha_1^*&= \frac{R^2 (R {-} 2) P - (4 R {-} 3)}{2(R{-}1)(3R^2{-}8R{+}9)} \,\epsilon - \frac{\mathrm{Q}_1(R,P)\epsilon^2}{16(R{-}1)^2(3R^2{-}8R{+}9)^3}+{\cal O}(\eps^3)\,,\\
\alpha_2^*&= -\frac{R(R^2{-}3R{+}3)P-(R{-}2)}{2 (R {-} 1) (3R^2{-}8R{+}9)}\,\epsilon + \frac{\mathrm{Q}_2(R,P)\epsilon^2}{16(R{-}1)^2(3R^2{-}8R{+}9)^3}+{\cal O}(\eps^3)\,,\\
\alpha_y^*&= \frac{R(R{-}3)P-(3R{-}1)}{4(R{-}1)(3R^2{-}8R{+}9)}\,\epsilon - \frac{\mathrm{Q}_y(R,P)\epsilon^2}{16(R{-}1)^2(3R^2{-}8R{+}9)^3}+{\cal O}(\eps^3)\,,
\end{aligned}
\eeq
with polynomials Q$_{1,2,y}(R,P)$ given in \eq{eq:Polynomials-GY12}. Similar (but lengthy) expressions for the scaling exponents are not given explicitly as they do not provide further insights. The explicit expressions make it evident that the conformal window will be modified due to higher-order effects, illustrated in Fig.~\ref{fig:Win_GY12_NNLO} 
for a selection of negative $\eps$ (left panel) and positive $\eps$ (right panel). In the left panel, starting with $\eps=0^-$, we observe that the red and yellow regions shrink with increasing $|\eps|$. On the other hand, we also observe that a new region (brown) is opening up that does not exist in the limit $\eps\to 0^-$. These fixed points are in accord with all basic constraints such as positivity, unitarity or the $a$-theorem, but since these regions originate from when $|P|$ is parametrically large, 
their reliability must be checked against higher-order corrections. 

\begin{figure}[t]
\includegraphics[width=.335\textwidth,valign=c]{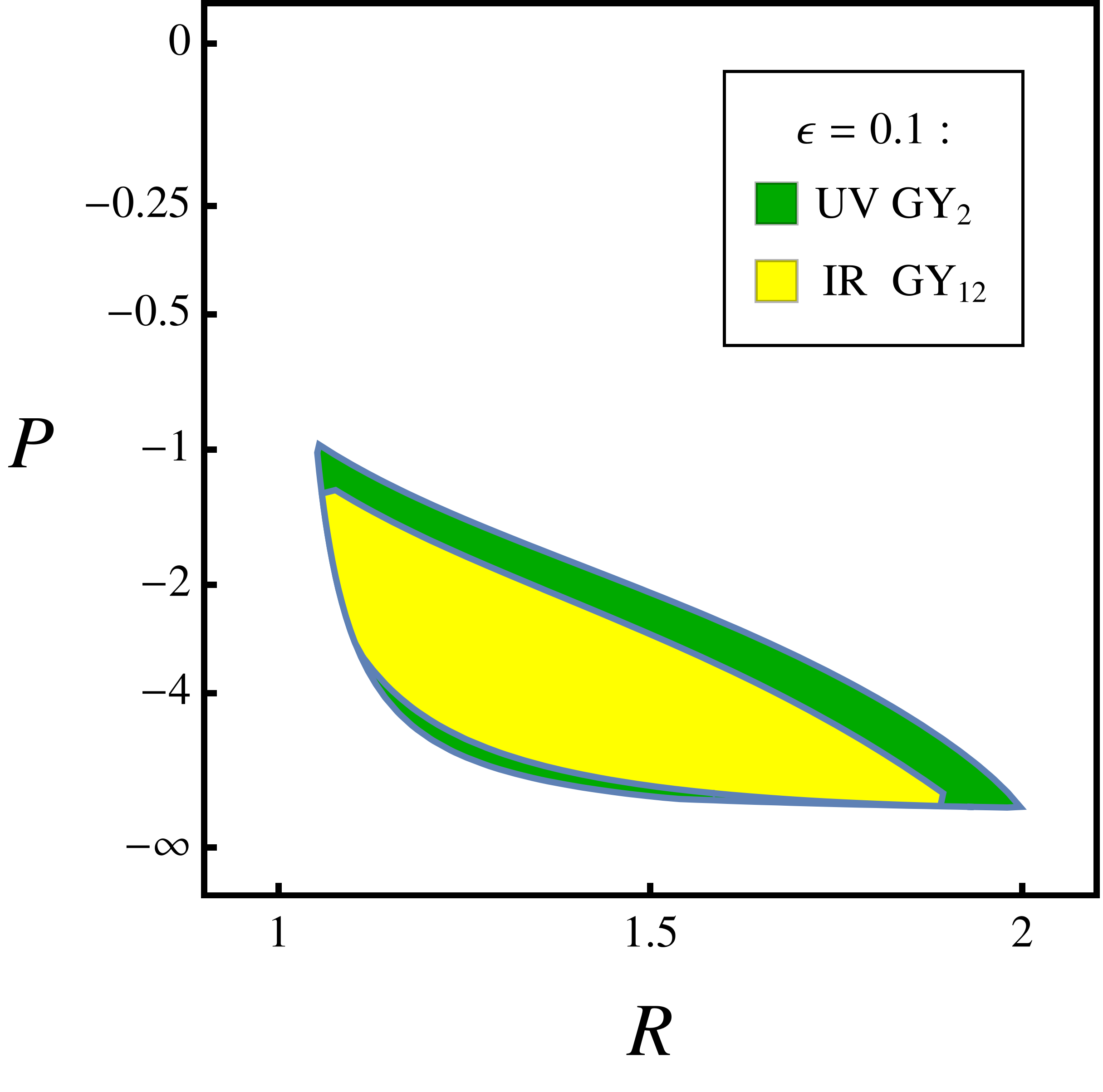}
\caption{Slices of the \gy{2} (green) and \gy{12} (yellow) conformal windows at NLO, shown exemplarily for $\eps=0.1$. Both regions exactly coincide for $\eps\to 0^+$.}
\label{fig:GY2_GY12_superposition}
\end{figure}

In Fig.~\ref{fig:GY2_GY12_superposition}, we compare the regions of existence for the fixed points \gy{2} and \gy{12}, exemplarily for $\epsilon=0.1$. We have already observed that their conformal windows agree provided $\epsilon\ll 1$, the reason being that \gy{12} disappears into the unphysical region by tunneling parametrically through \gy{2}. In doing so, the critical exponent at \gy{2} related to $B_{\rm eff, 1}$ changes sign, implying that \gy{2} ceases to be a UV fixed point. Consequently, the UV conformal window of \gy{2} coincides exactly with the IR conformal window of \gy{12}. Further, given that the fixed point structure is globally constrained by e.g.~Fig.~\ref{fig:FP-structure}, and that the physics does not change by increasing $\epsilon$, their boundaries must also coincide non-perturbatively, for any viable $\epsilon$. However, we observe from Fig.~\ref{fig:GY2_GY12_superposition} that boundaries do not agree. The mismatch should be taken as a measure of the approximation error due to using NLO perturbation theory at finite $\eps$. Provided $R>1$, the same discussion holds for the co-existence of the UV fixed point \gy{1} with the IR fixed point \gy{12}.

A new scenario arises in the regime with $R < 1$. Here, \gy{1} can be an ultraviolet fixed point, yet the IR fixed point \gy{12} never arises. It follows that RG trajectories emanating from \gy{1} invariably run towards a regime of strong coupling and confinement in the IR. This is very different from what happens for $R>1$, where trajectories that emanate from either \gy{1} or \gy{2} invariably run into the conformal IR sink \gy{12}. We emphasise that this new effect is not visible in the strictly perturbative regime $|\eps|\ll 1$ which, due to \eq{eq:positivity-constraints-reduced}, only probes the $R>1$ region.

\subsection{Higher-Order Effects for Model Building}

\begin{figure}[t]
\includegraphics[width=.8\textwidth,valign=c]{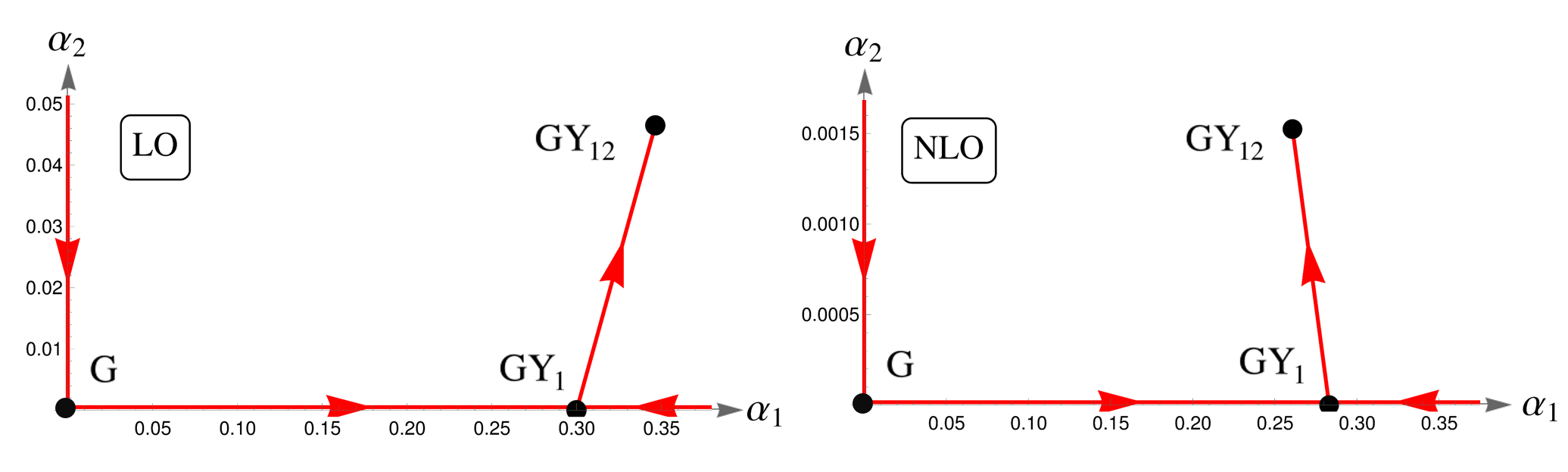}
\caption{Comparison of UV fixed points and the UV-IR connecting separatrix at LO (left panel) and NLO (right panel), illustrating that higher order effects can tilt the separatrix from $\alpha_1|_{\rm IR}> \alpha_1|_{\rm UV}$ to $\alpha_1|_{\rm IR}< \alpha_1|_{\rm UV}$, also mildly decreasing the value $\alpha^*_1|_{\rm UV}$. We also notice that $\alpha^*_2|_{\rm NLO}\ll \alpha^*_2|_{\rm LO}$ at GY${}_{12}$.}
\label{fig:bsm}
\end{figure}

We briefly discuss our results from the viewpoint of UV-safe supersymmetric extensions of the Standard Model. In \cite{Hiller:2022hgt}, a set of ${\cal O}(100)$ candidates for interacting UV fixed points has been identified to leading order in perturbation theory, based on $R$-parity violating extensions of the MSSM with new quark singlets and leptons and up to ten new superpotential couplings. In all settings, UV fixed points are qualitatively of the \gy{1} or \gy{2}-type, together with IR fixed points of the \gy{12}-type. Interestingly, UV-IR connecting trajectories in all models can be matched to the Standard Model (see Fig.~7 of \cite{Hiller:2022hgt}), with the sole but decisive caveat that the matching scale comes out too low, typically in the GeV energy range. It is important to emphasise that the mechanism for asymptotic safety does not leave room to tune the matching scale. Rather, it is entirely determined through the UV-IR connecting separatrix and the size of gauge couplings when coinciding with the RG trajectory of the Standard Model \cite{Hiller:2022hgt}.

It has been argued that the low matching scale could be an artefact of perturbation theory, and that higher-loop effects may enhance the matching scale by either decreasing the size of $\alpha^*_1|_{\rm UV}$, or by tilting the UV-IR connecting trajectory towards smaller values $\alpha^*_1|_{\rm IR}\ll \alpha^*_1|_{\rm UV}$. Interestingly, our NLO results indicate that both of these effects can happen, as illustrated in Fig.~\ref{fig:bsm}. While the hierarchy $\alpha^*_1|_{\rm IR}/\alpha^*_1|_{\rm UV} >1$ is hard-wired at LO (left panel), at NLO (right panel) we learn from \eqref{eq:NNLO-GY1}, \eqref{eq:NNLO-GY2}, and \eqref{eq:alphas-GY12-NNLO} that the hierarchy can indeed be inverted $\alpha^*_1|_{\rm IR}/\alpha^*_1|_{\rm UV}<1$ (in our models down to about $\alpha^*_1|_{\rm IR}/\alpha^*_1|_{\rm UV} \approx 0.8$). It follows that the running coupling $\alpha_1(\mu)$ can reach values below $\alpha^*_1|_{\rm UV}$ at NLO, and hence a higher matching scale than at LO \cite{Hiller:2022hgt}. In addition, we observe that the UV fixed point coupling at \gy{1} for $\eps<0$ becomes reduced (by about $10\%$ in our models), $\alpha^*_1|_{\rm NLO}<\alpha^*_1|_{\rm LO}$, see \eqref{eq:NNLO-GY1}, thus pulling into the same direction.\footnote{However, this is not the case for the UV fixed point \gy{2} (and $\eps>0)$ where higher-order corrections arise with the same sign, see \eqref{eq:NNLO-GY2}.} We conclude that it is worth revisiting the fixed point candidates of \cite{Hiller:2022hgt} at NLO, or even non-perturbatively \cite{Bond:2022xvr}.

\section{\bf Discussion}\label{sec:sum}

Combining exact methods from supersymmetry with perturbation theory and large-$N$, we have put forward a comprehensive analysis of conformal fixed points in general semi-simple supersymmetric gauge theories coupled to chiral superfields with or without a superpotential. Following up on \cite{Bond:2017lnq,Bond:2017suy,Bond:2022xvr}, we were particularly interested in new phenomena related to the semi-simple nature of the theory, and the availability of interacting UV fixed points that may serve as templates for model building. The sets of isolated fixed points (Fig~\ref{fig:FP-structure}) were investigated for general semi-simple gauge groups and in templates with unitary gauge groups. We determined scaling dimensions, phase diagrams, and conformal windows to leading order in a small Veneziano parameter $|\eps|\ll 1$ while keeping field multiplicities as free parameters (Figs.~\ref{fig:CW_BZ1_BZ2_NLO} - \ref{fig:CW_GY12_NLO}), and the ``phase space" of distinct quantum field theories (Fig.~\ref{fig:pFPall}). We further classified theories according to their UV behaviour and their spectra of isolated high- or low-energy conformal fixed points (Tab.~\ref{tab:pAll_AF} and~\ref{tab:pAll_AS}). Results include asymptotically free theories with a range of IR critical points (Figs.~\ref{fig:Flow-region-A},~\ref{fig:Flow-region-B},~\ref{fig:Flow-region-D}), asymptotically non-free theories that are nevertheless UV-complete and interacting both in the UV and the IR (Fig.~\ref{fig:Flow-region-F}), theories where one of the gauge sectors is both UV-free and IR-free yet interacting otherwise, and UV-incomplete effective theories that display IR conformal fixed points (Fig.~\ref{fig:Flow-eff}). The pattern of results is generic and not tied to the template models studied here. 

In order to find the conformal windows for finite Veneziano parameter $|\epsilon|\lesssim 1$, we extended the study to three-loop accuracy (Figs.~\ref{fig:Win_GY1_NNLO},~\ref{fig:GY2_NNLO_epsNeg},~\ref{fig:Win_GY12_NNLO}). Here, unitarity and the $a$-theorem turned out to be more constraining than for parametrically small Veneziano parameter. This is particularly relevant for theories with UV fixed points as illustrated in Figs.~\ref{fig:GY1_projections} and \ref{fig:GY2_projections}. We also observe that three-loop bounds on conformal windows (Figs.~\ref{fig:GY1_projections},~\ref{fig:comparison_Full_Loop_or_Eps}) are in good agreement with the infinite-order results of \cite{Bond:2017suy}, except for the strong-coupling boundaries of parameter space where $|\epsilon|$ is of order unity. Some of the findings at three loops suggest the existence of fixed points with new flow patterns (Fig.~\ref{fig:unphysical_relevancy}) that are strictly unavailable at weak coupling (Fig.~\ref{fig:relevancy_Gauge}). Closer inspection showed that the latter are incompatible with the $a$-theorem and must be discarded. In future work, it will be important to clarify whether other types of strongly-coupled UV fixed points may exist -- different from those established at weak coupling \cite{Bond:2016dvk,Bond:2017suy,Bond:2022xvr} or through Seiberg duality \cite{SEIBERG1995129,Barnes:2005zn}, yet in accord with all known constraints, e.g.~\cite{Martin:2000cr,Intriligator:2015xxa,Bajc:2017xwx}. 

Our results are also of interest for model building, the reason being that UV-completing asymptotically non-free supersymmetric theories via an interacting fixed point is more constraining, and more predictive, than without supersymmetry \cite{Bond:2022xvr}. For the MSSM, perturbative extensions with UV fixed points have been found \cite{Hiller:2022hgt}; however, they ultimately fail because the matching scale to Standard Model physics comes out too low \cite{Hiller:2022hgt}. Our results indicate that three-loop corrections may very well lower the critical coupling and tilt UV-IR connecting trajectories into the favoured direction (Fig.~\ref{fig:bsm}) to enhance the matching scale. It will then be interesting to revisit the models of \cite{Hiller:2022hgt} using improved approximations in perturbation theory and beyond. We look forward to coming back to this in the future.

\stoptoc
\section*{Acknowledgements}

We thank Gudrun Hiller for discussions. This work is supported by the \textit{Deutsche Akademische Austauschdienst} (DAAD) under the PRIME Fellowship, by the Science and Technology Facilities Council (STFC) under the Consolidated Grant No.~ST/X000796/1, and by a CERN Associateship.

\resumetoc

\newpage

\appendix

\renewcommand{\thesection}{{\bf \Alph{section}}}

\section{\bf Auxiliary Expressions}\label{app:beta-functions-NNLO}

In this Appendix, we provide results for RG beta functions up to three-loop order, anomalous dimensions, and auxiliary expressions exploited in the main text. We begin with the expressions for beta functions up to three-loop order, scaled according to the Veneziano limit. Using the general results of \cite{Einhorn:1982pp,Machacek:1983tz} we find the gauge beta functions up to three loops for our models as
\bea
	\beta_1^{(1)} &=& 
	\displaystyle
	2\alpha_1^2\, \eps 
\nonumber
\,,\\
	\beta_1 ^{(2)} &=& 
	\displaystyle
	2\alpha_1^2\big[(6+4\eps)\alpha_1 + 2R \alpha_2 
	- 4R(3 +\eps- R)\alpha_y\big]
\label{beta13}
\,,\\ \nonumber
	\beta_1^{(3)} &=& 4\alpha_1^2\left[2\eps\alpha_1^2 
			 - R\left(2\alpha_1\, \gam{\Psi}{1}+\gam{\Psi}{2}\right)
			 -(3 +\eps- R)\left(2\alpha_1\,\gam{\psi}{1} +\gam{\psi}{2}\right)\right]\,,
\eea
and		 
\bea\nonumber
	\beta_2^{(1)} &=& 
		\displaystyle
2\alpha_2^2\,P\eps
\,,\\
\label{beta23}
	\beta_2^{(2)} &=& 
		\displaystyle
2\alpha_2^2\left[(6+4P\eps)\alpha_2 + \02R\alpha_1 
		- \04R(3 - R+\eps)\alpha_y\right]
\,,\\ \nonumber
	\beta_2^{(3)} &=& 
	4\alpha_2^2\left[2 P \eps\alpha_2^2 -\frac1R \left(2\alpha_2\gam{\Psi}{1}+ \gam{\Psi}{2}\right)
\right.\\&&\nonumber 	\quad\quad \left. 
	- \frac{3-R+\eps}{R}\left( 2\alpha_2\,\gam{\chi}{1} +\gam{\chi}{2}\right)
- \left(4 + P \eps - \frac{4+\eps}{R}\right)\left( 2\alpha_2\,\gam{Q}{1}+\gam{Q}{2}\right)\right]\,,
\eea
where $\gamma_i^{(k)}$ is the anomalous dimension of the superfield $i$ in $k$-th loop accuracy. Moreover, the non-renormalisation of the superpotential dictates that the Yukawa beta function is given non-perturbatively by 
\bea
\label{betay}
\beta_y 
&= &2\alpha_y\big[\gamma_\psi+ \gamma_{\Psi}+ \gamma_{\chi}\big]\,,
\eea
valid for any loop order. In the perturbative analysis, the anomalous dimensions of chiral superfields are required up to two-loop accuracy,
\beq\label{gamma1}
\begin{array}{rcl}
	\gam{\psi}{1} &= &\displaystyle
	R\,\alpha_y - \alpha_1\,,\\[1ex]
	\gam{\psi}{2} &= &\displaystyle-R\,\alpha_y\left(\gam{\Psi}{1} + \gam{\chi}{1}\right) - \alpha_1 \gam{\psi}{1} + 4\,\eps\,\alpha_1^2\,,\\[2ex]
	\gam{\Psi}{1} &=&\displaystyle (3 - R+\eps)\alpha_y - \alpha_1 -\alpha_2\,,\\[1ex]
	\gam{\Psi}{2} &=&\displaystyle -(3 - R+\eps)\,\alpha_y\left(\gam{\psi}{1} + \gam{\chi}{1}\right) - (\alpha_1 + \alpha_2)\gam{\Psi}{1}
				+ 4\,\eps\,\alpha_1^2 + 4 \,P\,\eps\,\alpha_2^2\,,\\[2ex]
	\gam{\chi}{1} &=&\displaystyle \alpha_y - \alpha_2\,,\\[1ex]
	\gam{\chi}{2} &= &\displaystyle-\alpha_y\left(\gam{\psi}{1} + \gam{\Psi}{1}\right) - \alpha_2\gam{\chi}{1}
				+ 4\, P\,\eps\,\alpha_2^2\,,\\[2ex]
	\gam{Q}{1} &= &\displaystyle-\alpha_2\,,\\[1ex]
	\gam{Q}{2} &= &\displaystyle-\alpha_2\, \gam{Q}{1} + 4 \,P\,\eps \,\alpha_2^2\,.
\end{array}
\eeq
With these expressions at hand, one extracts the leading and subleading terms in $\eps$ of fixed point couplings and universal scaling exponents, and the size of conformal windows.

The analytic results for the critical exponents of the fully interacting \bz{12} fixed point are given in \eqref{eq:NLO-thetas-BZ12}. They involve the following polynomials
\beq\label{eq:polynomials-thetas-BZ12}
\begin{aligned}
Q_1^{\mathrm{BZ}_{12}}(R,P) &= 3R^2(R^2+9)P^2 - 18R(R^2+1)P + 3(9R^2+1)\,,\\
Q_2^{\mathrm{BZ}_{12}}(R,P) &=\, 9R^4(R^4{-}14R^2{+}81)P^4 {-} 12R^3(3R^2{-}4R{-}9)(3R^2{+}4R{-}9)P^3
\\ &\quad
 {+} 2R^2(243R^4 {-} 826R^2{+}243)P^2
 {-} 12R(9R^2{-}4R{-}3)(9R^2{+}4R{-}3)P
 \\ &\quad
 {+} 9(81R^4{-}14R^2{+}1)\,.
\end{aligned}
\eeq
Similarly, the first subleading (in $\eps$) contributions to couplings at the \gy{12} fixed point as given in \eqref{eq:alphas-GY12-NNLO} 
involve the polynomials
\begin{equation}\label{eq:Polynomials-GY12}
\begin{aligned}
Q_1(R,P)=&[R^3(R{-}2)(27R^4{-}155R^3{+}425R^2{-}573R{+}324)]P^2
\\& {-}[2R^2(9R^5{-}105R^4{+}385R^3{-}605R^2{+}350R{-}18)]P
\\&{+}(27R^6{+}63R^5{-}1053R^4{+}3505R^3{-}5178R^2{+}3624R{+}972)\,,\\
Q_2(R,P)=&[R^2(36R^6{-}314R^5{+}1250R^4{-}2853R^3{+}3909R^2{-}3033R{+}1053)]P^2
\\& 
{-}[2R(6R^5{-}60R^4{+} 277R^3{-}653R^2{+}765R{-}351)]P
\\& 
{+}(18R^5{-}210R^4{+}787R^3{-}1383R^2{+}1123R{-}351)\,,\\
Q_y(R,P)=&[R^2(18R^5{-}139R^4{+}487R^3{-}963R^2{+}1035R{-}486)]P^2
\\& 
{+}[2R(9R^5{-}24R^4{-}23R^3{+}127R^2{-}78R{-}27)]P
\\& 
{+}(108R^5{-}657R^4{+}1683R^3{-}2087R^2{+}1221R{-}252)\,.
\end{aligned}
\end{equation}
Lastly, in Sec.~\ref{sec:CW}, we discussed how chiral superfield anomalous dimensions impact upon unitarity, see Figs.~\ref{fig:comparison_Full_Loop_or_Eps} and~\ref{fig:comparison_gamma_3}. Here, we provide the relevant expressions for $\gamma_\psi$. The exact infinite order result has been derived in \cite{Bond:2022xvr} using $a$-maximisation and is given by
\begin{equation}
\begin{aligned}
	\gamma_\psi&= \frac{R\big[(2R{-}3{-}\eps)^2+3-\Delta\big]+\eps(2R{-}3{-}\eps)(2R{-}3{-}\eps+1)}{2(2R{-}3{-}\eps)\big[(2R{-}3{-}\eps)^2-(3{+}\eps)\big]}\,,
\end{aligned}
\end{equation}
with $\Delta$ the positive root of $[(2R{-}3{-}\eps)^2+3]^2 + 8\eps(2R{-}3{-}\eps)^2$. The result can be expanded as a power series in $\eps$, see \eq{gamma_expansion}, and the first two coefficients are in full agreement with the direct results from perturbation theory, as they must. The exact coefficient at cubic order reads
\begin{equation}\label{eq:comparing-gamma-3-gy1}
\hspace{-.1cm}
\begin{aligned}
\left.\gamma_\psi^{(3)} \right|_{\rm exact} &= \frac{-8 R^7{+}108 R^6{-}618 R^5{+}1948 R^4{-}3642 R^3{+}4014 R^2{-}2385 R{+}576}{128(R^2-3 R+3)^5}\,.
\end{aligned}
\end{equation}
On the other hand, simultaneously solving the three-loop gauge beta function and the two-loop expressions for $\gamma_\psi$, and expanding the result in powers of $\eps$, we find the exact linear and quadratic coefficients together with an infinite set of incomplete higher order coefficients. The first of these, the incomplete cubic coefficient, reads
\begin{equation}\label{eq:incomplete}
\begin{aligned}
\left.\gamma_\psi^{(3)} \right|_{\rm NLO}&= \frac{4 R^5{+}20 R^4{-}204 R^3{+}599 R^2{-}792 R{+}432}{64(R^2-3 R+3)^4}\,.\\
\end{aligned}
\end{equation}
The difference between \eq{eq:comparing-gamma-3-gy1} and \eq{eq:incomplete}, displayed in Fig.~\ref{fig:comparison_gamma_3}, relates to the (missing) four-loop gauge and the three-loop Yukawa terms which contribute at NNLO order.

\section{\bf Banks--Zaks Beyond Leading Order}
\label{app:BZs}

In this appendix, we summarise findings for Banks--Zaks fixed points beyond the leading order, following Sec.~\ref{sec:CW}.

At NLO accuracy, the \bz{1} fixed point continues to be physical $(\alpha>0$) provided $\eps<0$. The corrections to the fixed point value and effective one-loop coefficient are
\begin{equation}
\alpha_1^*=-\s016\eps + \s0{1}{12}\eps^2 + \mathcal{O}(\eps^3)\,, \quad B_{2;\eff}= \left(-2P + \s0{2}{3R}\right)\,\eps - \s0{2}{9 R}\,\eps^2 + \mathcal{O}(\eps^3)\,,
\end{equation}
and, to the critical exponents,
\begin{equation}
\vartheta_1= \s013\,\eps^2 - \s0{2}{9}\,\eps^3 > 0\,, \quad \vartheta_3= \s023\,\eps - \s0{2}{9}\,\eps^2 < 0\,,
\end{equation}
with definite signs for $\eps<0$. $\alpha_2$ is marginally relevant or marginally irrelevant depending on the sign of $B_{2;\eff}$. The NLO conformal windows and the relevancy of $\alpha_2$ are shown in Fig.~\ref{fig:Win_BZ1_NNLO} for three different values of $\eps$. As in the \gy{1} case, the conformal windows changes smoothly as $\eps$ grows.
\begin{figure}
\includegraphics[width=0.45\linewidth]{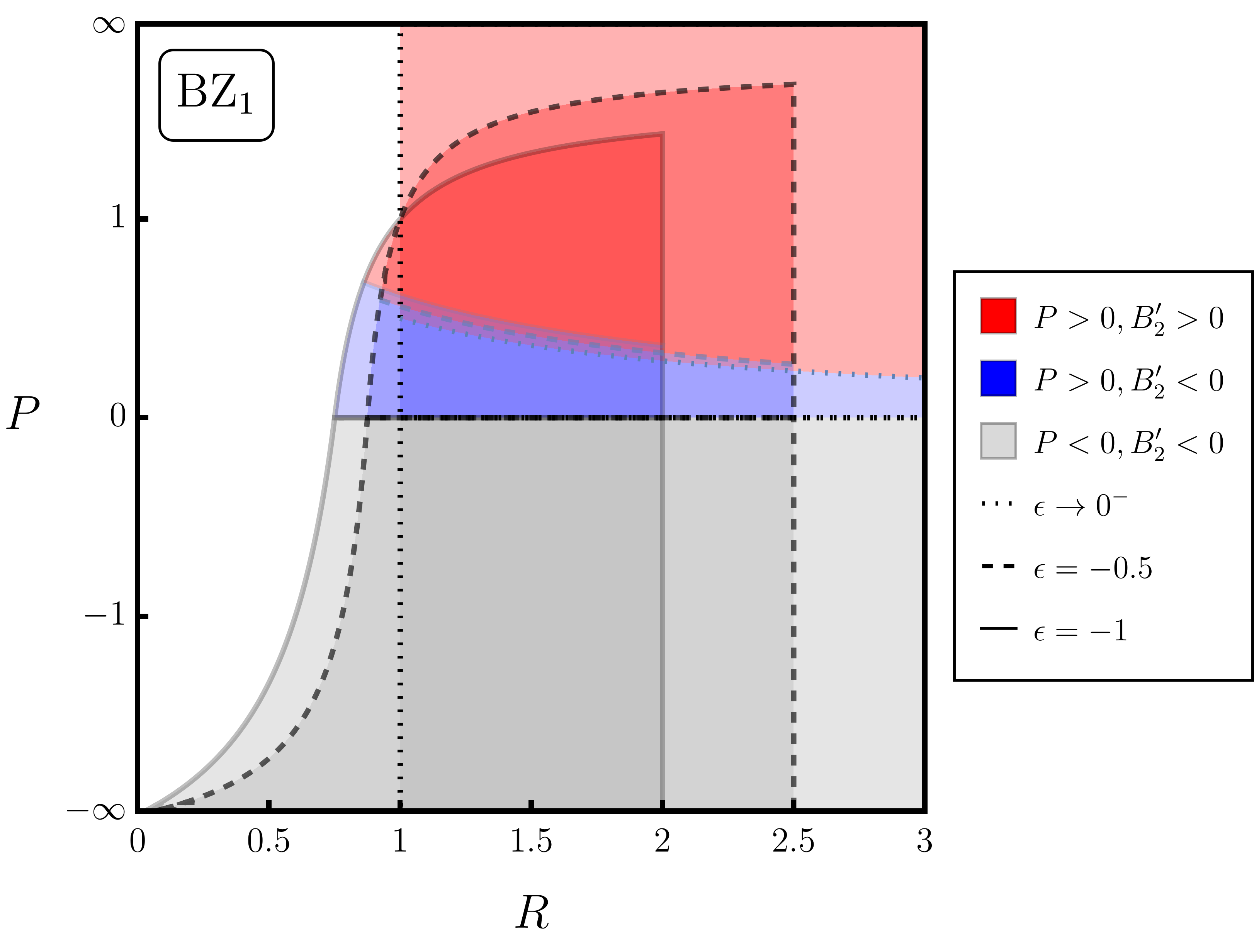}
\caption{\bz{1} conformal window at NLO for various $\eps<0$.} 
\label{fig:Win_BZ1_NNLO}
\end{figure}

For the \bz{2} fixed point, the NLO corrections read
\begin{equation}
\alpha_2^*= -\s016P\,\eps + \s0{1}{12}P^2\eps^2 + \mathcal{O}(\eps^3)\,, \quad B_{1;\eff}= \left( -2 + \s023 RP \right)\,\eps -\s0{2R}{9}P^2\eps^2 + \mathcal{O}(\eps^3)\,,
\end{equation}
and
\begin{equation}
\vartheta_2 = \s013 P^2\eps^2 - \s029 P^3\eps^3\,, \quad \vartheta_3 = \s023 P\eps - \s029 P^2\eps^2 \,,
\end{equation}
with $\alpha_1$ being marginally relevant or marginally irrelevant depending on the sign of $B_{1;\eff}$.

The new conformal windows is illustrated in Fig.~\ref{fig:Win_BZ2_NNLO}. As in the \gy{2} analysis, the results from the three-loop contributions can be divided into smooth deformations of the previously obtained windows with $P\eps<0$ and new conformal windows apparently opening up for $P\eps>0$. However, as in the \gy{2} case, such regions would imply in the existence of trajectories from the \bz{2} in the UV to the Gaussian in the IR and are then constrained by \eqref{eq:constraint-a-theorem}. It is easy to check, even analytically in this case, that such a condition is never satisfied, therefore, the new regions are unphysical, and we are left with only the deformations of the regions previously obtained in two-loop accuracy.
\begin{figure}[t]
\includegraphics[width=0.75\textwidth]{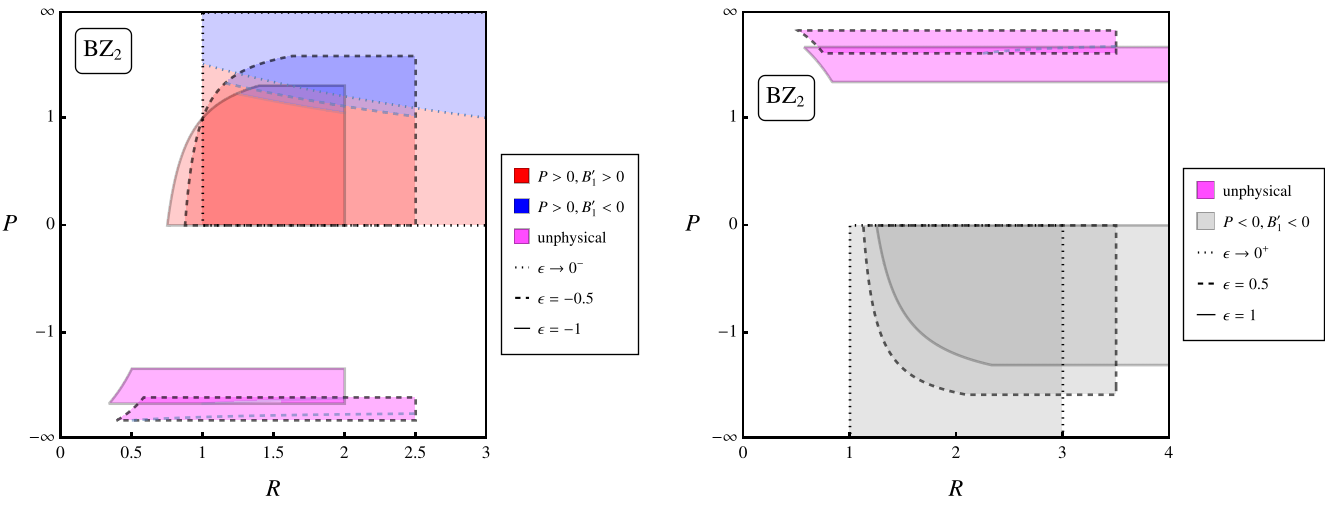}
\caption{\bz{2} conformal windows at NLO for various
$\eps<0$ (left panel) and $\eps>0$ (right panel).} \label{fig:Win_BZ2_NNLO}
\end{figure}

Finally, the fully interacting \bz{12} fixed point is still only physical for negative values of $\eps$, with the three-loop order contributions to the couplings at the fixed point being
\beq
\begin{aligned}
\alpha_1^*& = \s01{16}{(R P - 3)}\,\eps - \frac{1}{512}\left( \frac{R(5R{+}21)}{2}P^2 + (9R{+}1)P - \frac{33R{+}1}{2R} \right)\eps^2 + \mathcal{O}(\eps^3)\,, \\
\alpha_2^*& = -\s03{16}(P-\s01{3R})\,\eps + \frac{1}{512} \left( \frac{3(R{+}33)}{2}P^2 - \frac{(R{+}9)}{R}P - \frac{(21R{+}5)}{2R^2} \right)\eps^2 + \mathcal{O}(\eps^3)\,.
\end{aligned}
\eeq
The conformal windows within $P>0$, shown in Fig.~\ref{fig:Win_BZ12_NNLO}, can be viewed as a smooth deformation of the two-loop order result. However, for finite $\eps$, we observe a new region opening up for large values of $P\eps>0$ and close to the region of unphysical fixed points observed in Fig.~\ref{fig:Win_BZ2_NNLO}. These types of solutions cannot arise in settings that are under strict perturbative control. We therefore consider these solutions as spurious and outside the domain of validity of our approximations. Still, this parameter region would benefit form an all-order study using $a$-maximisation.
\begin{figure}
\includegraphics[width=0.38\linewidth]{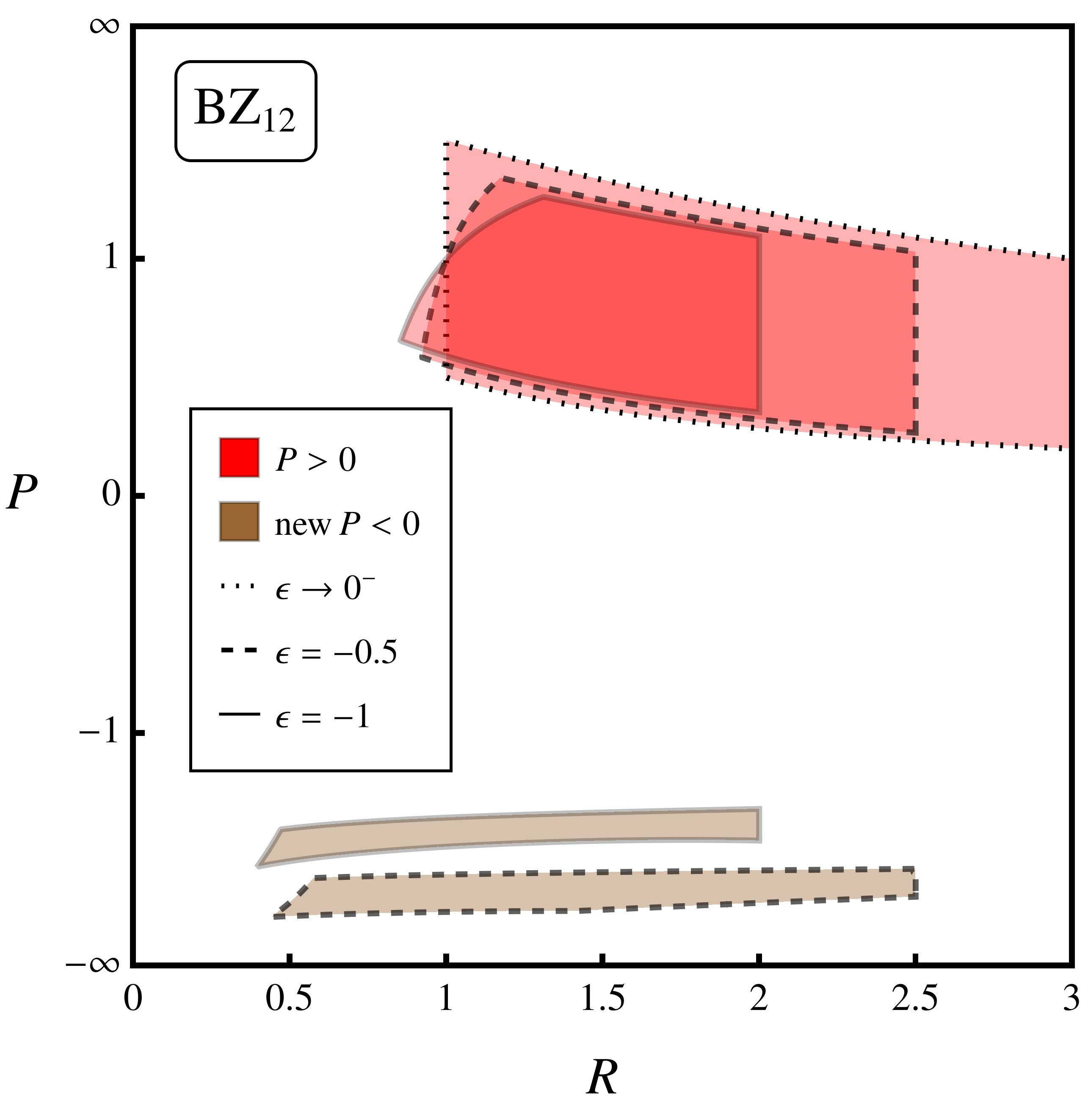}
\caption{\bz{12} conformal window at NLO for various $\eps<0$.}
\label{fig:Win_BZ12_NNLO}
\end{figure}

\bibliographystyle{JHEP}
\bibliography{bib_semi2,bib_New}

\end{document}